\documentclass[a4paper,11pt]{article}
\pdfoutput=1 

\usepackage{jheppub} 

\usepackage[T1]{fontenc} 

\usepackage{float, array, xspace, amscd, amsthm}
\usepackage{fancyhdr}
\usepackage{longtable}
\usepackage{comment}

\newtheorem{theorem}{Theorem}

\newtheorem{defn}{Definition}

\usepackage{amsmath}
\usepackage{amssymb}
\usepackage{mathtools}
\numberwithin{equation}{section}
\usepackage{hyperref}
\usepackage{color}

\newcommand{\ind}{\text{ind}}
\newcommand{\oth}{\text{otherwise}}

\newcommand{\be}{\begin{equation}}
\newcommand{\ee}{\end{equation}}
\def\bea#1\eea{\begin{align}#1\end{align}}

\title{Line bundle cohomologies on CICYs \\ with Picard number two}
\author[a]{Magdalena Larfors,}
\author[a]{Robin Schneider}

\affiliation[a]{Department of Physics and Astronomy,Uppsala University\\ SE-751 20 Uppsala, Sweden}

\emailAdd{magdalena.larfors@physics.uu.se, robin.schneider@physics.uu.se}

\abstract{We analyse line bundle cohomologies on all favourable co-dimension two Complete Intersection Calabi Yau (CICY) manifolds of Picard number two. Our results provide further evidence that the cohomology dimensions of such line bundles are given by analytic expressions, which change between regions in the line bundle charge space. This agrees with  recent observations of CY line bundles presented in Refs \cite{Constantin:2018hvl,Klaewer:2018sfl}. In many cases, the expressions for bundle cohomology dimensions are  polynomial functions of the line bundle charges (of degree at most 3), and the regions are cones.  A more novel observation is that for some CICY manifolds,  the cohomologies are more succinctly determined by recursive relationships. There can also be  boundaries between regions where a polynomial fit fails, and we link these exceptional cases to irregular behaviour of the index of the line bundle.
Finally, our observations provide evidence for similarities in the line bundle cohomologies for CICY manifolds that share rows in the configuration matrix. Among such related CICY manifolds, we find both that the line bundle charge space is partitioned in the same manner, and that the same, or closely related, analytical descriptions apply for the cohomology dimensions in these regions.}

\preprint{UUITP-18/19}

\begin{document}

\maketitle
\flushbottom

\newpage


	\section{Introduction}

	Line bundles play an important role in string model building. In the framework of heterotic compactifications on a Calabi Yau (CY) manifold, they make up the main building blocks for constructing vector bundles. This is a very fruitful patch of the string landscape and has resulted in more than $63000$ unique realistic standard models \cite{Anderson:2011ns,Anderson:2012yf,Anderson:2013xka}.  Line bundles are also  important ingredients in the construction of vector bundles that can stabilise the geometric moduli in heterotic compactifications \cite{Anderson:2010mh,Anderson:2011cza}. 
	
	Finding a realistic compactification requires the calculation of the dimensions of line bundle cohomologies; these dimensions determines the number of fermion generations, and the dependence of the cohomology dimension on geometric moduli carries information about moduli stabilisation.  The computation of this topological information relies on algebraic geometry, as we will review in section \ref{sec:2} (see also Refs \cite{Hubsch:1992nu,Anderson:2008ex}), and involves a non-trivial analysis of exact sequences, which can be implemented on a computer.  In the case of Complete Intersection Calabi Yau (CICY) manifolds \cite{Candelas:1987kf}, which will be the manifolds we consider in this paper, line bundle cohomologies have been algorithmically implement in Mathematica in \cite{CICYpackage}, and more recently in Python \cite{CICYtoolkit} based on \cite{Anderson:2008ex}.
	For toric ambient spaces there exists an efficient algorithm for line bundle cohomologies  \cite{Rahn:2010fm,CohomOfLineBundles:Algorithm,cohomCalg:Implementation}. In the general case, these algorithmic algebraic geometry programs requires significant computational resources.
	
	Recently, it has been observed that the cohomology dimensions follow some remarkable simple descriptions in the  space of line bundle charges \cite{Buchbinder:2013dna}. The charge space is divided into different regions, where in each region, the dimension is given by a polynomial in the line bundle charges of degree less or equal to the dimension $n$ of the involved CY $n$-fold. This observation seems natural, since the cohomology dimension is constrained by the index of the line bundle, which in turn can be calculated via integration over the Chern classes and hence expressed as a polynomial of degree $n$. As a follow up to that observation analytic expressions have been derived for CY hypersurfaces in toric ambient spaces \cite{Klaewer:2018sfl}, and for CICYs \cite{Constantin:2018hvl}. In this paper we  continue the analysis of the latter, for two reasons. First, the set of CICY manifolds has long been shown to be well suited for string compactifications as these manifolds contain many freely acting symmetries needed for model building \cite{Braun:2010vc}. Second, the extension of systematic studies as in \cite{Anderson:2011ns,Anderson:2012yf,Anderson:2013xka} to manifolds with $h^{1,1} > 7$ is currently, in part, bottlenecked by the complexity of the computations of the involved line bundle cohomologies. However, these manifolds are expected to have significantly more realistic standard models than the ones analysed so far \cite{Constantin:2018xkj}. Analytic control over line bundle cohomologies would  allow us to bypass the computationally expensive algebro-geometric analysis, and improve our ability to construct realistic string vacua. 
	
	In Ref \cite{Constantin:2018hvl} the authors find analytic equations for mainly codimension one CICYs with Picard number one to four. The line bundle charge space splits into regions, and the cohomology dimensions in each region is described by a polynomial of degree 3. The majority of these regions follow a cone structure also observed in Ref \cite{Klaewer:2018sfl}. The regions are often separated by boundaries, which follow their own polynomial descriptions. In the analysis  of Ref \cite{Constantin:2018hvl}, charges in the range $m_i \in \{-10, ... , 10\}$ have been generated and the polynomial description holds for all of these points, even though the amount of data generated is magnitudes higher than needed for the polynomial fits. The analysis also contains analytic descriptions for line bundle cohomologies on quotients by freely acting symmetries of Calabi Yau manifolds.
	
	In this paper we extend the analysis of Ref \cite{Constantin:2018hvl} to seven additional CICY manifolds, which all have Picard number and co-dimension two.  In section \ref{sec:2} we will briefly recapitulate the algebro-geometric computation of the line bundle cohomologies in terms of maps between ambient space cohomologies. Section \ref{sec:3} contains the main results of this paper. Here, we present analytic expressions for the eight codimension 2 CICYs with $h^{1,1} = 2$. We analyse these results, and discuss similarities among analytic expressions for line bundle cohomologies of different manifold  in section \ref{sec:4}. We summarise our results and provide an outlook to future studies in section \ref{sec:5}.

	
	\section{Line bundle cohomologies on CICY manifolds from Leray tableaux}
	
	\label{sec:2}
	
	In this section we will briefly introduce our notation and recapitulate the theory of  Complete Intersection Calabi Yau (CICY) manifolds, which were first studied in Ref \cite{Candelas:1987kf}. This section is provided for completeness and can be skipped by the expert reader.
	
	\begin{defn}
		A \textbf{CICY} is a set of hypersurfaces in products of complex projective space such that the resulting manifold $X$ is Calabi Yau. We write for $X$ the following configuration matrix:
		\begin{align}
		X =  \left[
		\begin{array}{c||ccc}
		n_0 & q^0_1 & \cdots & q^0_{K} \\
		\vdots & \vdots & \ddots & \vdots \\
		n_r & q^{r}_1 & \cdots & q^{r}_K  \\
		\end{array}
		\right]^{h^{1,1},h^{2,1}}_{\chi}
		\end{align}
		Here, $n_i$ denotes the degree of the $i$-th projective space and $q^i_j$ is the degree of the $j$-th hypersurface in the $i$-th projective space. We denote the Hodge numbers and Euler characteristic of $X$ by $h^{1,1},h^{2,1}$ and  $\chi$, respectively.
	\end{defn}
	
	A line bundle $L$ over $X$ is denoted by $L = \mathcal{O}_X (m_0, ..., m_r)$, where $m_i$ are the charges in front of the K\"ahler forms $J_i$ corresponding to each $\mathbb{P}^{n_i}$. The starting point for determining the line bundle cohomology dimensions $h^\bullet(X, L)$ of this line bundle is the Koszul resolution
	\begin{align}
	\label{eq: Koszul}
	0 \rightarrow \mathcal{L} \otimes \wedge^K \mathcal{N}^*_X \rightarrow \mathcal{L} \otimes \wedge^{K-1} \mathcal{N}^*_X \rightarrow \dots \rightarrow \mathcal{L} \otimes \mathcal{N}^*_X \rightarrow \mathcal{L} \rightarrow L \rightarrow 0.
	\end{align}
	Note, that $L$ is the restriction of the ambient space line bundle $\mathcal{L} = \mathcal{O}_\mathcal{A}(m_0,, ..., m_r)$ to $X$. This long exact sequence can be split into several short exact sequences, to which we can each associate a long exact sequence in cohomology. In the end this procedure makes it possible to express $H(X,L)$ in terms of $H(\mathcal{A}, \mathcal{L})$, $H(\mathcal{A}, \mathcal{L} \otimes \mathcal{N}^*_X)$, and ambient space cohomologies valued in direct products of $\mathcal{L}$ with higher wedge products of $\mathcal{N}^*_X$. These ambient space cohomologies are determined by applying the Bott--Borel--Weil theorem \cite{Hubsch:1992nu,Anderson:2008ex}.
	\begin{theorem} \label{th:bbw}
		Let $\mathbb{F}$ be a flag space and $\mathcal{V} \sim  (a_1, \dots , a_{n_1} | \dots | d_1,\dots,d_{n_f})$ a holomorphic homogeneous vector bundle over $\mathbb{F}$. Then:
		\begin{enumerate}
			\item Homogeneous vector bundles $\mathcal{V}$ over $\mathbb{F}$ are in 1-1 correspondence with the $U(n_1)\times \dots \times U(n_F)$ representations.
			\item The cohomology $H^i(\mathcal{A},\mathcal{V})$ is non-zero for \textbf{at most one value} of $i$, in which case it provides an irreducible representation of $U(N)$, $H^i(\mathbb{F},\mathcal{V}) \approx (c_1,\dots,c_N)\mathbb{C}^N$.
			\item The bundle $(a_1, \dots , a_{n_1} |\dots|d_1,\dots,d_{n_f})$, determines the cohomology group $(c_1, \dots ,c_N)$, according to the following algorithm:
			\subitem 1) Add the sequence $1,\dots,N$ to the entries in $(a_1, \dots , a_{n_1} 
			|\dots|$ $d_1,\dots,d_{n_F})$.
			\subitem 2) If any two entries in the result of step 1 are equal, all cohomology vanishes; otherwise proceed.
			\subitem 3) swap the minimum number $(=i)$ of neighbouring entries required to produce a strictly increasing sequence.
			\subitem 4) Subtract the sequence $1,\dots,N$ from the result of previous step, to obtain $(c_1, \dots , c_N)$.
		\end{enumerate}
	\end{theorem}
	It is clear from 2) that the majority of ambient space cohomologies are going to vanish. The whole procedure of splitting the Koszul resolution and expressing $H(X,L)$ in terms of ambient space cohomologies is summarised in a Leray tableaux.
	\begin{defn}
		A Leray tableaux is given by
		\begin{align}
		E_{i+1}^{j,k} = \frac{\text{Ker}(d_i : E_{i}^{j,k} (\mathcal{L}) \rightarrow E_i^{j+i-1,k-1}(\mathcal{L}) )}{\text{Im}(d_i : E_{i}^{j+i-1,k+i} (\mathcal{L}) \rightarrow E_i^{j,k}(\mathcal{L}) )}
		\end{align}
		where the first term is defined as
		\begin{align}
		E_1^{j,k} (\mathcal{L}) := H^j (\mathcal{A},\mathcal{L} \otimes \wedge^K \mathcal{N}^*_X ), \quad k=0,\dots,K; \; j=0,\dots, \text{dim}(\mathcal{L}).
		\end{align}
		This is a complex defined by differential maps $d_i : E_i^{j,k} \rightarrow E_i^{j-1,k-1}$ for $j=1,2,\dots$ going to infinity.
	\end{defn}
	Now it is well known that for line bundles this complex terminates already at $i=2$ \cite{Hubsch:1992nu}. The cohomology dimensions are then given by
	\begin{align}
	\label{eq: hql}
h^q (X, L) = \sum_{m=0}^{K} \text{rk}E_2^{q+m,m} (\mathcal{L}).
	\end{align}
	Determining the cohomology dimensions, thus requires us to compute the rank of the differential maps $d_i$ between different ambient space cohomologies, which can be implemented on a computer  \cite{CICYpackage,CICYtoolkit,Rahn:2010fm,CohomOfLineBundles:Algorithm,cohomCalg:Implementation}. These computations are generally computationally very expensive, as one has to find all linearly independent relationships in an immense linear system of equations.	The details of this analysis go beyond the scope of the paper, and we refer the interested reader to the very thorough analysis \cite{Anderson:2008ex} from which we have taken our notation (see also \cite{CICYpackage}) or the classic reference \cite{Hubsch:1992nu}.
	
	While the calculation of cohomology dimensions can be fairly complicated, there are a few key theorems that  make our life easier. We will summarise these theorems here, and again refer the reader to \cite{Constantin:2018hvl,Anderson:2008ex,Hubsch:1992nu,huybrechts2006complex}, and references therein, for more details. The most important result is Serre duality relating cohomologies of different degree. For a Calabi Yau it implies
	\begin{align}
		H^{n-q}(X,L) = H^q(X,L^*)
	\end{align}
	where $n$ is the dimension of the Calabi Yau.
	
	Second in line is the Kodaira vanishing theorem which states that on a CY manifold $X$ 
	\begin{align}
	H^q(X,L) = 0 \text{ for } q>0 \text{ if } L \text{ is positive,}
	\end{align}
	i.e. $\forall i, m_i > 0 $. This theorem greatly simplifies calculations, as combined with Serre duality it directly applies to $m_0, m_1 < 0$ and $m_0, m_1 > 0$, which is almost half of the line bundle charge space that we will explore.
	
	Moreover, recall that the index of the line bundle can, by the Atiyah--Singer index theorem,  be computed from integrating the Chern class. For CY threefolds, we have
	\begin{align}
	\label{eq: ind}
	\ind(L) = \sum_{q=0}^{n} (-1)^q h^q(X,L) = \frac{1}{6} d_{rst} m^r m^s m^t + \frac{1}{12} c_2^r m_r
	\end{align}
	where $d_{rst}$ are the triple intersection numbers  and $c_2^r$ is the $r$-th component of the vectorised second Chern class of the CY threefold $X$, and $m_r$ are the line bundle charges. When combined with the Kodaira vanishing theorem, we see we may determine all cohomology dimensions $h^q(X,L)$ without computing a single Leray tableaux map, whenever the line bundle charges are all positive or negative. From equation \eqref{eq: ind} we also directly observe, that the index is given by a polynomial of degree 3 with only cubic and linear terms in the charges.

	Finally, we mention a slightly modified result of a vanishing theorem due to Kobayashi \cite{kobayashi1987differential}. We state it here as presented in \cite{Anderson:2012yf}. The cohomologies $H^0$ and $H^3$ of poly stable vector bundles with slope zero vanish. For line bundles this translates to the fact that whenever
	\begin{align}
		\mu(L) = c_1^i d_{ijk} t^j t^k \stackrel{!}{=} 0
	\end{align}
	somewhere in the K\"ahler cone ($t_j, t_k > 0$) we have $h^0(X,L) = 0 =h^3(X,L)$.

	
	\section{Formulae for line bundle cohomologies on example CICY manifolds}
	
	\label{sec:3}

	In this paper, we study all favourable CICY manifolds of codimension two with Picard number two. There are exactly eight manifolds satisfying these conditions, corresponding to  number 7806, 7833, 7844, 7858, 7882, 7883, 7885 and 7888 in the standard CICY list \cite{Candelas:1987kf}, and one of these spaces has been explored in \cite{Constantin:2018hvl}. A benefit of working with these manifolds is that their line bundles are determined by two charges, which means that our results can easily be presented graphically. Furthermore, it is possible to create sufficient data for the analysis without running in to computational limitations. We have computed the line bundle cohomology dimensions for  charges in the range $m_i \in \{-10 , ... , 10\}$. In the event that this  provided too few data points  to make the polynomial fit converge, additional data points were computed to find and confirm our analytical descriptions. A python implementation of our line bundle cohomology calculations based on \cite{Anderson:2008ex} can be found here \cite{CICYtoolkit}.

	In the rest of this section we will present our results for these manifolds, in a slightly changed order. We begin with the manifolds 7806, 7882 and 7888 which all share $[1||02]$ as a first entry in their configuration matrix. The next in line are 7833, 7844 and 7883 with $[2||21]$ and finally 7858 and 7885 with $[1||11]$. The reason for this classification is that the analytical expressions for the line bundle cohomologies in each group share certain features, as will become more clear in the discussion section. Since Serre duality relates the dimensions  $h^0$,$h^3$,  as well as $h^1$,$h^2$, we will only present results for $h^0$ and $h^1$. 	A common feature of all these CICYs is that the charge space splits into five regions where different expressions for the cohomology dimensions apply. In the rest of this section, we will see that this partition can in part be observed already from the structure of the Leray tableaux, whereas other features remain obscure on this level of the analysis, and requires computation of the maps in this tableaux.

	\subsection{Ambient space and Leray tableaux}
	
	\label{sequence}

	As mentioned in section \ref{sec:2} combining the Kodaira vanishing theorem, Serre duality and the index we can fully determine the first $(m_0 > 0, m_1 > 0)$ and third quadrant $(m_0 <0, m_1 < 0)$ of $h^0, h^1$ without considering any maps. Hence, we will only discuss the second $(m_0 <0, m_1 > 0)$ and fourth $(m_0 > 0, m_1 < 0)$ quadrant here.
	From \eqref{eq: hql} we find
	\begin{align}
	\label{eq: hql0}
		h^0 &= \text{rk} E_2^{0,0} + \text{rk} E_2^{1,1} + \text{rk} E_2^{2,2}\\
	\label{eq: hql1}
	h^1 &= \text{rk} E_2^{1,0} + \text{rk} E_2^{2,1} + \text{rk} E_2^{3,2}\\
	\label{eq: hql2}
	h^2 &= \text{rk} E_2^{2,0} + \text{rk} E_2^{3,1} + \text{rk} E_2^{4,2}\\
	\label{eq: hql3}
	h^3 &= \text{rk} E_2^{3,0} + \text{rk} E_2^{4,1} + \text{rk} E_2^{5,2},
	\end{align}
	where we recall that $E_i^{j,k}$ are entries in the Leray tableaux. The rank of these entries are determined by the Bott--Borel--Weil theorem, and are highly ambient space dependent. Thus, it is useful to split our discussion in two parts, first for the ambient space $\mathcal{A} = \mathbb{P}^1 \times  \mathbb{P}^4$ and second for $\mathcal{A} = \mathbb{P}^2 \times \mathbb{P}^3$.
	
	\subsubsection{$\mathcal{A} = \mathbb{P}^1 \times  \mathbb{P}^4$}
	
	\label{sec:311}
	
	When calculating the ambient space cohomology, we find that in the second quadrant $(m_0 < 0, m_1 > 0)$ only $H^1 (\mathcal{A}, \mathcal{V})$, where $\mathcal{V} = \mathcal{L} \otimes \wedge^k \mathcal{N}^*$ with $k=0,1,2$, is non vanishing. This follows directly from the Bott--Borel--Weil theorem. In particular, the algorithm presented in theorem \ref{th:bbw}, shows that a negative first and positive second charge requires a flip  in the first projective space, and thus a swap of dim$(\mathbb{P}^1) = 1$ entries so that the only non trivial ambient space cohomologies occur for $j=1+0=1$. Similarly, a negative second and positive first charge, implies a non trivial ambient space cohomology at $j=4$.
	Going through the Leray tableaux we find that only $E^{1,1}_2$ and $E^{1,0}_2$ are non trivial. Thus from \eqref{eq: hql0} and \eqref{eq: hql1} follows 
	\begin{align}
	\label{eq: 142h0}
		h^0(X,L) &= \text{rk} \big( \text{Ker} ( H^1( \mathcal{A}, \mathcal{L} \otimes \mathcal{N}^*) \rightarrow H^1( \mathcal{A}, \mathcal{L}) ) \big) \nonumber \\
		& \qquad - \text{rk} \big( \text{Im} ( H^1( \mathcal{A}, \mathcal{L} \otimes \wedge^2 \mathcal{N}^*) \rightarrow H^1( \mathcal{A}, \mathcal{L} \otimes \mathcal{N}^*) ) \big) 	\end{align}
	and
	\begin{align}
		\label{eq: 142h1}
		h^1 (X,L) &= h^1(\mathcal{A}, \mathcal{L}) - \text{rk} \big( \text{Im} ( H^1( \mathcal{A}, \mathcal{L} \otimes \mathcal{N}^*) \rightarrow H^1( \mathcal{A}, \mathcal{L}) ) \big).
	\end{align}
	Further, $h^2(X,L) = 0 = h^3(X,L)$. 
	
	Meanwhile when considering the fourth quadrant $(m_0 > 0, m_1 < 0)$ only $H^4 (\mathcal{A}, \mathcal{V})$ is non vanishing. Thus the following relations hold $h^0(X,L) = 0 = h^1(X,L)$ and by equations \eqref{eq: hql2} and \eqref{eq: hql3} we find
	\begin{align}
		\label{eq: 144h2}
		h^2 (X,L) = \text{rk} \big( \text{Ker} ( H^4( \mathcal{A}, \mathcal{L} \otimes \wedge^2 \mathcal{N}^*) \rightarrow H^4( \mathcal{A}, \mathcal{L} \otimes \mathcal{N}^*) ) \big) 
	\end{align}
	and
	\begin{align}
		\label{eq: 144h3}
		h^3 (X, L) &= \text{rk} \big( \text{Ker} ( H^4( \mathcal{A}, \mathcal{L} \otimes \mathcal{N}^*) \rightarrow H^4( \mathcal{A}, \mathcal{L}) ) \big) \nonumber \\
		& \qquad - \text{rk} \big( \text{Im} ( H^4( \mathcal{A}, \mathcal{L} \otimes \wedge^2 \mathcal{N}^*) \rightarrow H^4( \mathcal{A}, \mathcal{L} \otimes \mathcal{N}^*) ) \big) .
	\end{align}
	
	\subsubsection{$\mathcal{A} = \mathbb{P}^2 \times  \mathbb{P}^3$}
	
	\label{sec:312}

	For the ambient space $\mathcal{A} = \mathbb{P}^2 \times  \mathbb{P}^3$ we have non trivial ambient space cohomology in the second quadrant for $H^2(\mathcal{A}, \mathcal{V})$. Hence only $E^{2,2}_2, E^{2,1}_2$ and $E^{2,0}_2$ are non trivial such that
	\begin{align}
		\label{eq: 232h0}
		h^0 (X,L) &= \text{rk}  \big(\text{Ker} (H^2(\mathcal{A}, \mathcal{L} \otimes \wedge^2 \mathcal{N}^*) \rightarrow H^2(\mathcal{A}, \mathcal{L} \otimes \mathcal{N}^*)) \big) 
	\end{align}
	and
	\begin{align}
	\label{eq: 232h1}
	h^1 (X, L) &= \text{rk}\big(\text{Ker} (H^2(\mathcal{A}, \mathcal{L} \otimes \mathcal{N}^*) \rightarrow H^2(\mathcal{A}, \mathcal{L}) \big) \nonumber \\
	& \qquad - \text{rk} \big( \text{Im}(H^2(\mathcal{A}, \mathcal{L} \otimes \wedge^2 \mathcal{N}^*) \rightarrow H^2(\mathcal{A}, \mathcal{L} \otimes \mathcal{N}^*)) \big)
	\end{align}
	and
	\begin{align}
	\label{eq: 232h2}
	h^2 (X, L) &= h^2 (\mathcal{A}, \mathcal{L}) - \text{rk} \big( \text{Im} (H^2(\mathcal{A}, \mathcal{L} \otimes \mathcal{N}^*) \rightarrow H^2(\mathcal{A}, \mathcal{L} )) \big) 
	\end{align}
	and $h^3(X,L) = 0$.
	On the other hand the only non vanishing ambient space cohomology in the fourth quadrant is $H^3 (\mathcal{A}, \mathcal{V})$. From which follows, $h^0(X,L) = 0$,
	\begin{align}
	\label{eq: 234h1}
	h^1 (X, L) &= \text{rk}\big(\text{Ker} (H^3(\mathcal{A}, \mathcal{L} \otimes \wedge^2 \mathcal{N}^*) \rightarrow H^3(\mathcal{A}, \mathcal{L} \otimes \mathcal{N}^*)) \big) 
	\end{align}
	and
	\begin{align}
	\label{eq: 234h2}
	h^2 (X, L) &= \text{rk} \big( \text{Ker} (H^3(\mathcal{A}, \mathcal{L} \otimes \mathcal{N}^*) \rightarrow H^3(\mathcal{A}, \mathcal{L} ) \big) \nonumber \\
	& \qquad - \text{rk} \big( \text{Im}(H^3(\mathcal{A}, \mathcal{L} \otimes \wedge^2 \mathcal{N}^*) \rightarrow H^3(\mathcal{A}, \mathcal{L} \otimes \mathcal{N}^*)) \big)
	\end{align}
	and
	\begin{align}
	\label{eq: 234h3}
	h^3 (X, L) &= h^{3}(\mathcal{A}, \mathcal{L}) - \text{rk} \big( \text{Im} (H^3(\mathcal{A}, \mathcal{L} \otimes \mathcal{N}^*) \rightarrow H^3(\mathcal{A}, \mathcal{L} )) \big) .
	\end{align}

	\subsection{CICY manifolds with $[1||02]$}
	\label{sec:102}
	In this, and the following sections, we present the cohomology dimensions for line bundles, and the analytical expressions that describe this data. Some care is needed here, as the expressions we provide can only be shown to be correct for line bundle charges in the range $(-10,...,10)$, and we will comment more on this in section \ref{sec:4}. With this cautionary remark in mind, we will present the results graphically in order to clearly show the regions where different analytical descriptions apply. For the favourable codimension two CICY manifolds with Picard number two, there are five regions. We colour code these regions using purple, yellow, red, blue and green. A purple region indicates that the cohomology vanishes, and a red region indicates that the cohomology dimension is given by the index of the line bundle. In the yellow, blue and green regions the dimensions are given by other analytical expressions, that we determine for each manifold.
	
	We start with CICY manifolds that have configuration matrices that contain $[1||02]$, where we find that the cohomology dimensions always admits polynomial descriptions.

	\subsubsection{7806}
	
\begin{figure}
			\centering
			\begin{minipage}{0.5\textwidth}
				\centering
				\includegraphics[width=1.1\textwidth]{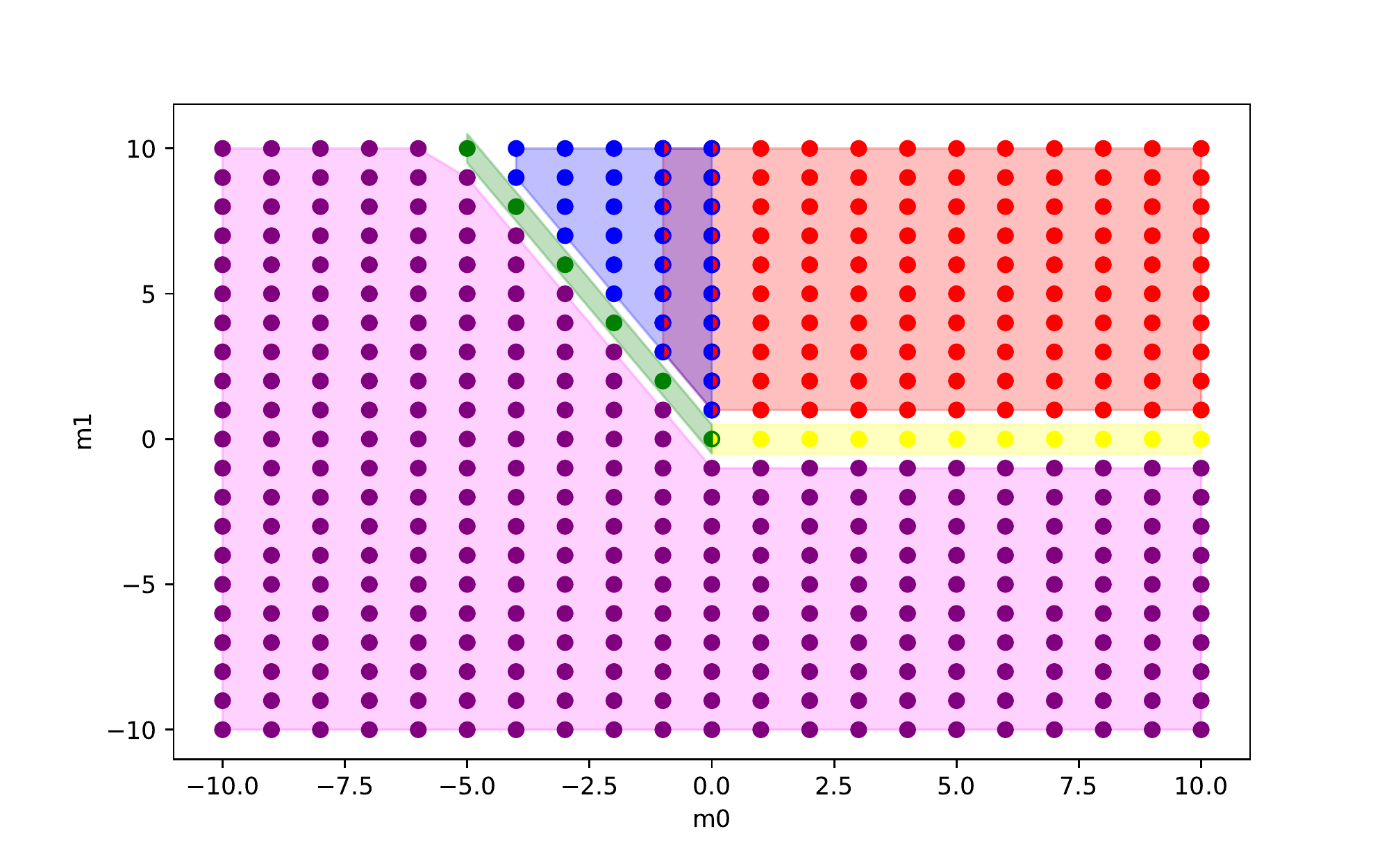} 
			\end{minipage}\hfill
			\begin{minipage}{0.5\textwidth}
				\centering
				\includegraphics[width=1.1\textwidth]{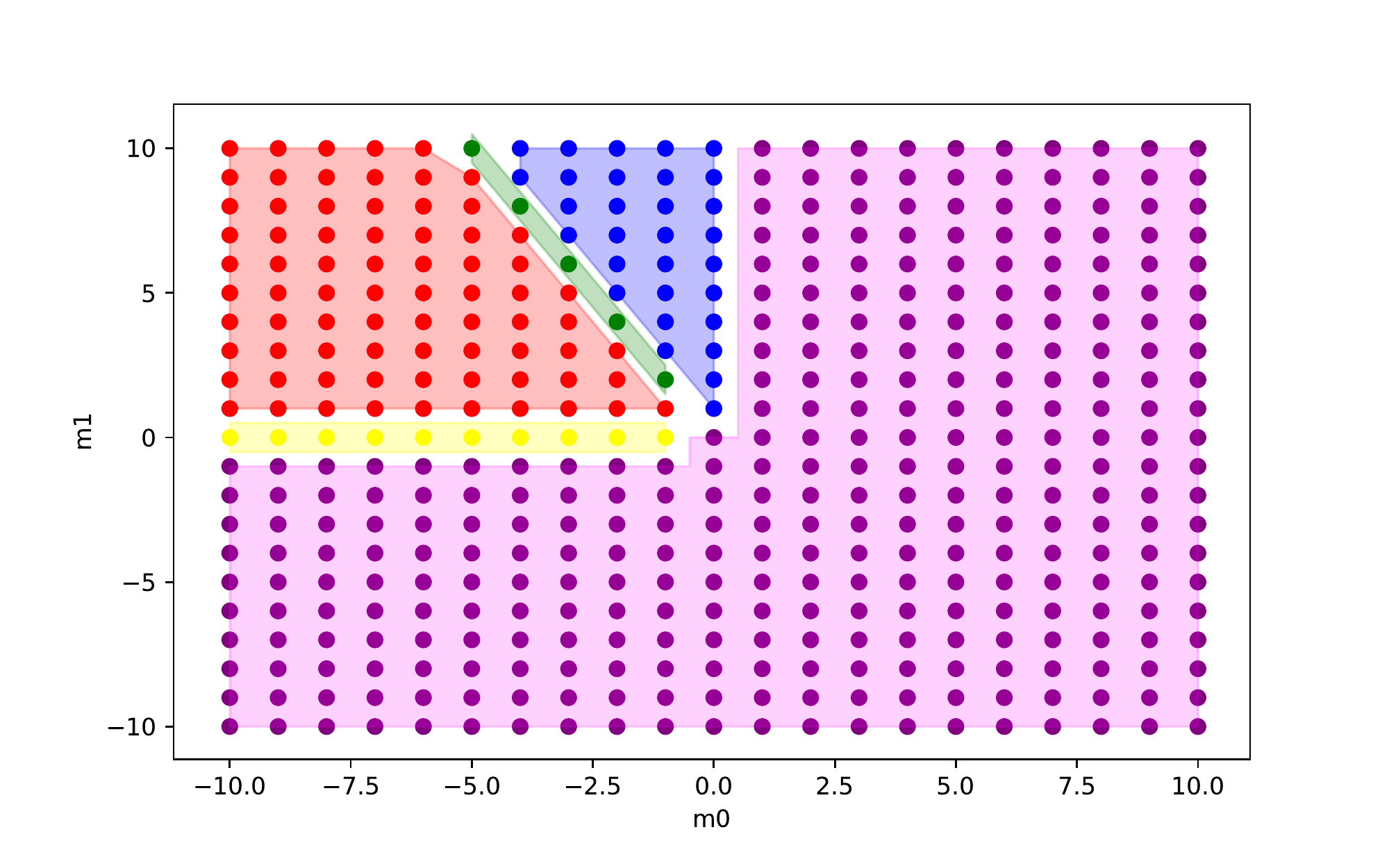} 
			\end{minipage}
			\caption{$h^0$ and $h^1$ for line bundles with charges $m_0$ and $m_1$ over the CICY \# 7806. There are five regions, in purple, yellow, red, blue and green, where the colors indicate  polynomial description found in \eqref{eq: 7806h0} and \eqref{eq: 7806h1}. Overlapping cones imply that there are two polynomial descriptions for the data.}
			\label{plt: 7806b} 
		\end{figure}
	
	Consider the manifold
	\begin{align}
	M_{7806} =  \left[
	\begin{array}{c||cc}
	1 & 0 & 2  \\
	4 & 3  & 2 
	\end{array}
	\right]^{2,56}_{-108}.
	\end{align}
	Its index $\ind (L)$ and slope $\mu(L)$ are given by
	\begin{align}
	\label{eq: 7806i}
	\ind (L) &= 3m_0m_1^2+2m_0+m_1^3+4m_1 \; ,\\
	\label{sl: 7806}
		\mu(L) &= 6 (m_0 + m_1) t_1^2 + 12 m_1 t_0 t_1.
	\end{align}
	The Leray tableaux computation of the cohomology dimensions for line bundle charges in $[-10,10]$ are consistent with the polynomial expressions
	\begin{align}
	\label{eq: 7806h0}
	h^0(X, L) = \begin{cases}
	1+m_0, & m_0 \geq 0, m_1 = 0 \\
	1-m_0, & m_0 \leq 0, m_1 = -2m_0 \\
	4m_0(1-m_0^2)+\ind(L), & m_0 \leq 0, m_1 > -2m_0 \\
	\ind(L), & m_0 > 0, m_1 > 0 \\
	0 &\oth 
	\end{cases}
	\end{align}
	and
	\begin{align}
	\label{eq: 7806h1}
	h^1(X, L) = \begin{cases}
	-1-m_0, & m_0 < 0, m_1 = 0 \\
	-\ind(L)+1-m_0, & m_0 < 0, m_1 = -2m_0 \\
	4m_0(1-m_0^2), & m_0 \leq 0, m_1 > -2m_0 \\
	-\ind(L), & m_0 < 0, 0 < m_1 < -2m_0 \\
	0 &\oth \; .
	\end{cases}
	\end{align}

	In figure \ref{plt: 7806b} we present this result graphically.  We note that the polynomial expressions is set by the  Kodaira vanishing theorem in the interior of the first quadrant, where only $h^0$ is non-vanishing, and hence $h^0 = \ind(L)$. On the boundary between the first and the fourth quadrant ($m_0 \geq 0, m_1=0$), both $h^0$ and $h^1$ are non-zero. In the second quadrant, there is a region between the lines $m_1 = 0$ and $m_1=-2m_0$ where both $h^0$ and $h^1$ are non-vanishing, and are described by polynomials of degree three. Beyond this region, for $m_1<-2m_0$, only $h^1$ is non-vanishing, and is given by the index. This implies that ranks of the kernel and image in \eqref{eq: 142h0} cancel exactly.

	By Serre duality, this behaviour is repeated for $h^2$ and $h^3$, but with a reflection about the origin:  the only non-vanishing cohomology in the third quadrant is $H^3$, and the fourth quadrant is divided into one region where both $h^2$ and $h^3$ are non-vanishing, and one region where only $h^2$ is non-vanishing.

	\subsubsection{7882}

	Consider the manifold
	\begin{align}
	M_{7882} =  \left[
	\begin{array}{c||cc}
	1 & 0 & 2  \\
	4 & 2  & 3 
	\end{array}
	\right]^{2,76}_{-148}.
	\end{align}
	Its index is given by
	\begin{align}
	\ind (L) = 3 m_0 m_1^2 +2 m_0+ \frac{2}{3} m_1^3 + \frac{25}{6} m_1
	\end{align}
	and the slope is
	\begin{align}
		\mu(L) = ( 6 m_0 + 4 m_1) t_1^2 + 12 m_1 t_0 t_1 .
	\end{align}
	Analysing the data yields
	\begin{align}
	\label{eq: 7882h0}
	h^0(X, L) = \begin{cases}
	1+m_0, & m_0 \geq 0, m_1 = 0 \\
	1-m_0, & m_0 \leq 0, m_1 = -3m_0 \\
	9m_0(1-m_0^2)+\ind(L), & m_0 \leq  0, m_1 > -3m_0 \\
	\ind(L), & m_0 > 0, m_1 > 0 \\
	0 &\oth 
	\end{cases}
	\end{align}
	and
	\begin{align}
	\label{eq: 7882h1}
	h^1(X, L) = \begin{cases}
	-1-m_0, & m_0 < 0, m_1 = 0 \\
	-\ind(L)+1-m_0, & m_0 < 0, m_1 = -3m_0 \\
	9m_0(1-m_0^2), & m_0 \leq 0, m_1 > -3m_0 \\
	-\ind(L), & m_0 < 0, 0 < m_1 < -3m_0 \\
	0 &\oth \; .
	\end{cases}
	\end{align}
	
\begin{figure}
		\centering
		\begin{minipage}{0.5\textwidth}
			\centering
			\includegraphics[width=1.1\textwidth]{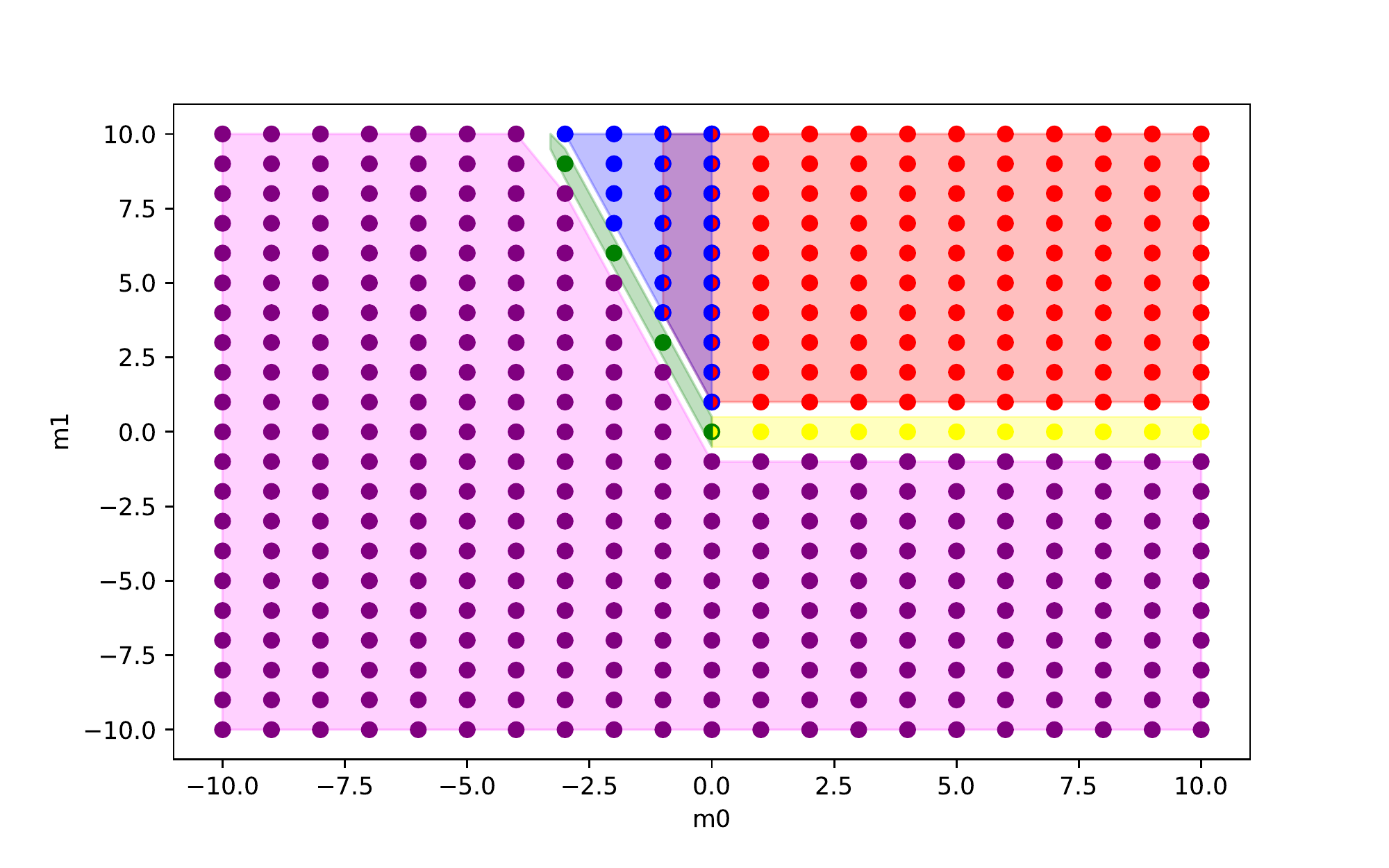} 
		\end{minipage}\hfill
		\begin{minipage}{0.5\textwidth}
			\centering
			\includegraphics[width=1.1\textwidth]{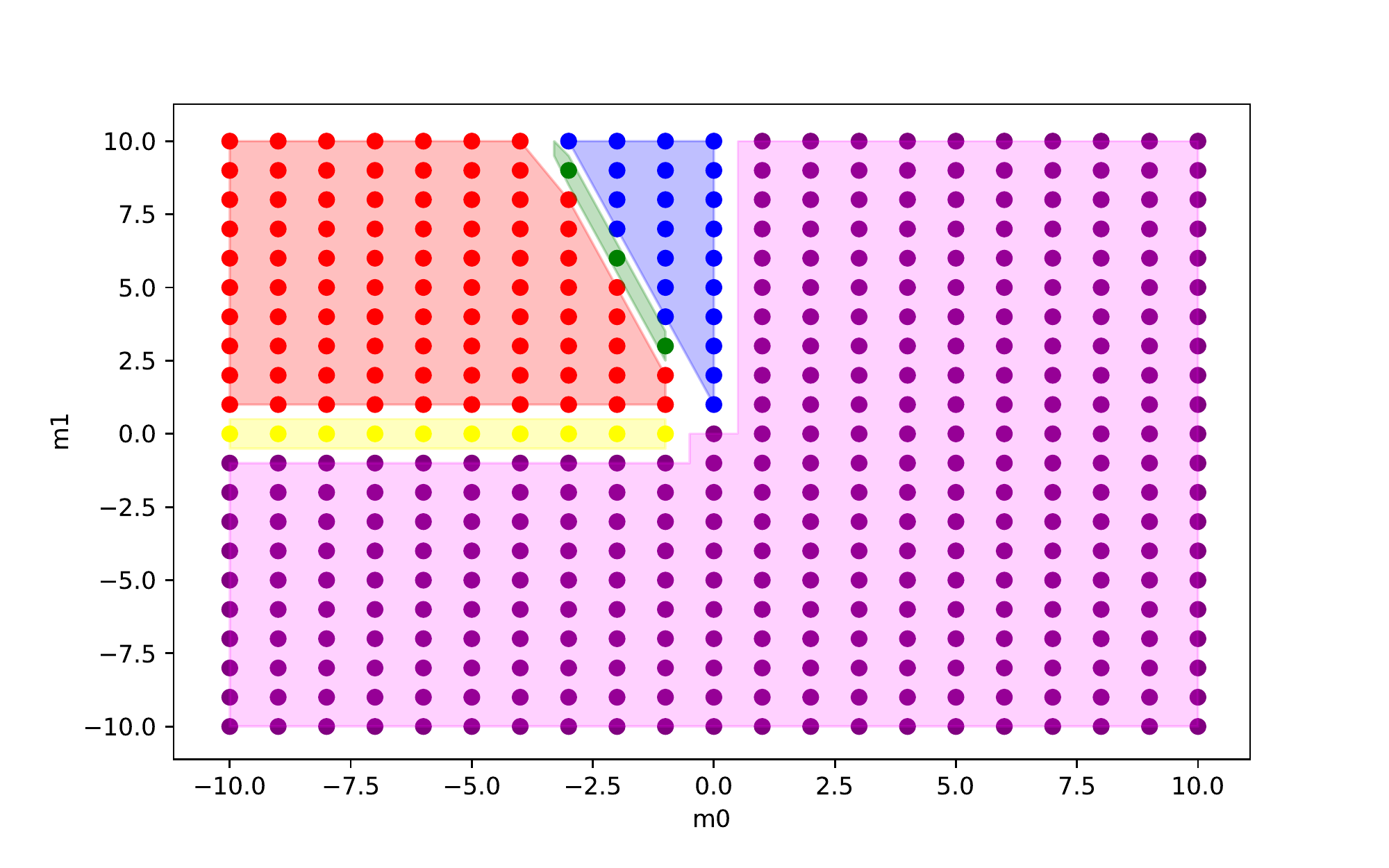} 
		\end{minipage}
		\caption{$h^0$ and $h^1$ for line bundles with charges $m_0$ and $m_1$ over the CICY \# 7882. Colors indicate regions with different  polynomial description found in \eqref{eq: 7882h0} and \eqref{eq: 7882h1}.  Near the $m_1$ axis, there are several polynomial descriptions $h^0$, which leads to an overlap between the blue and the red cone, and the green and the yellow cone.} 
		\label{plt: 7882b}
		\end{figure}

	There is a clear similarity between both the polynomial expressions and the partition of the line bundle charge space between this example and the previous one. This is particularly striking from figure \ref{plt: 7882b}, where we present the result of this subsection graphically. 
	
	\subsubsection{7888}

	\begin{figure}
		\centering
		\begin{minipage}{0.5\textwidth}
			\centering
			\includegraphics[width=1.1\textwidth]{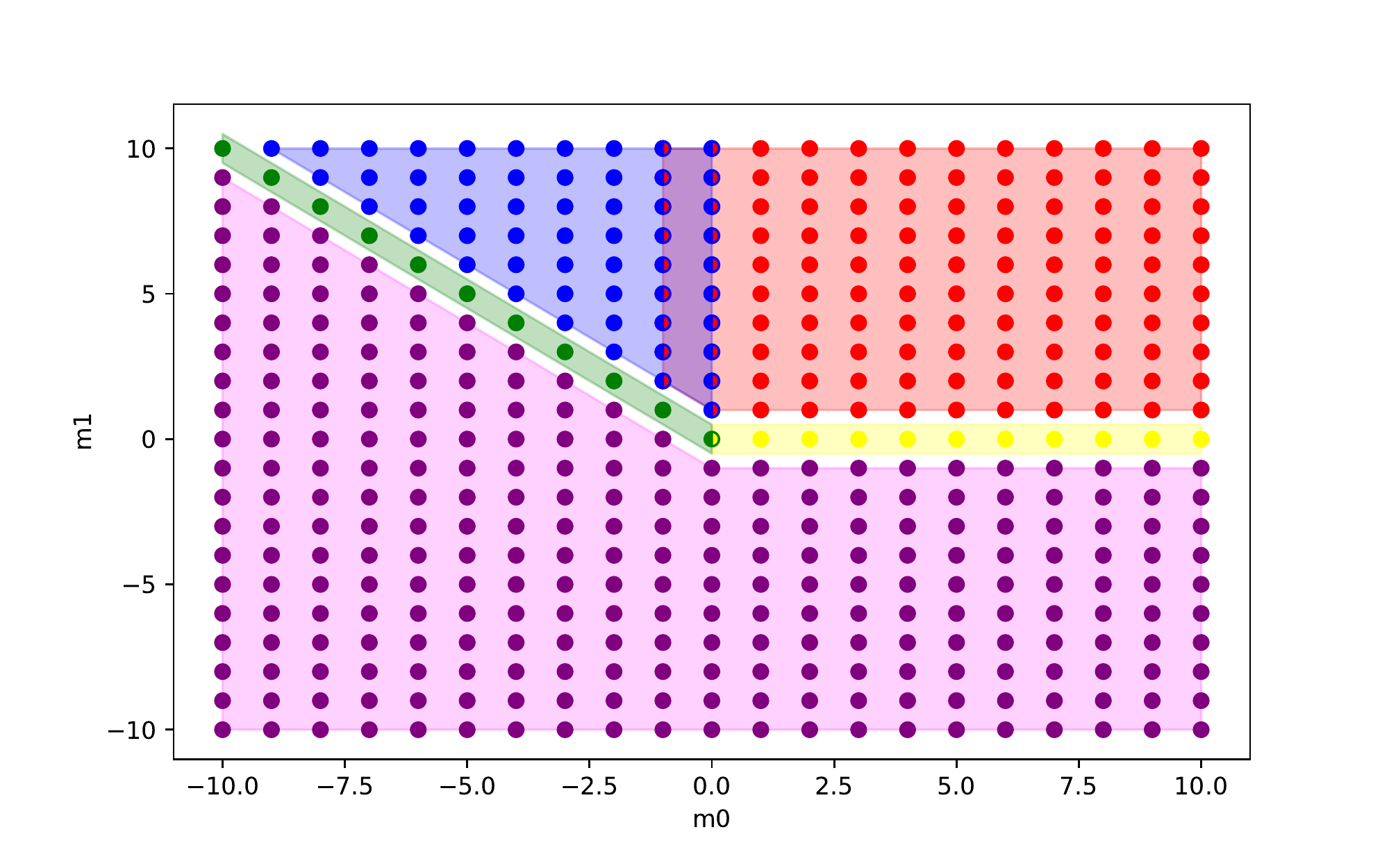} 
		\end{minipage}\hfill
		\begin{minipage}{0.5\textwidth}
			\centering
			\includegraphics[width=1.1\textwidth]{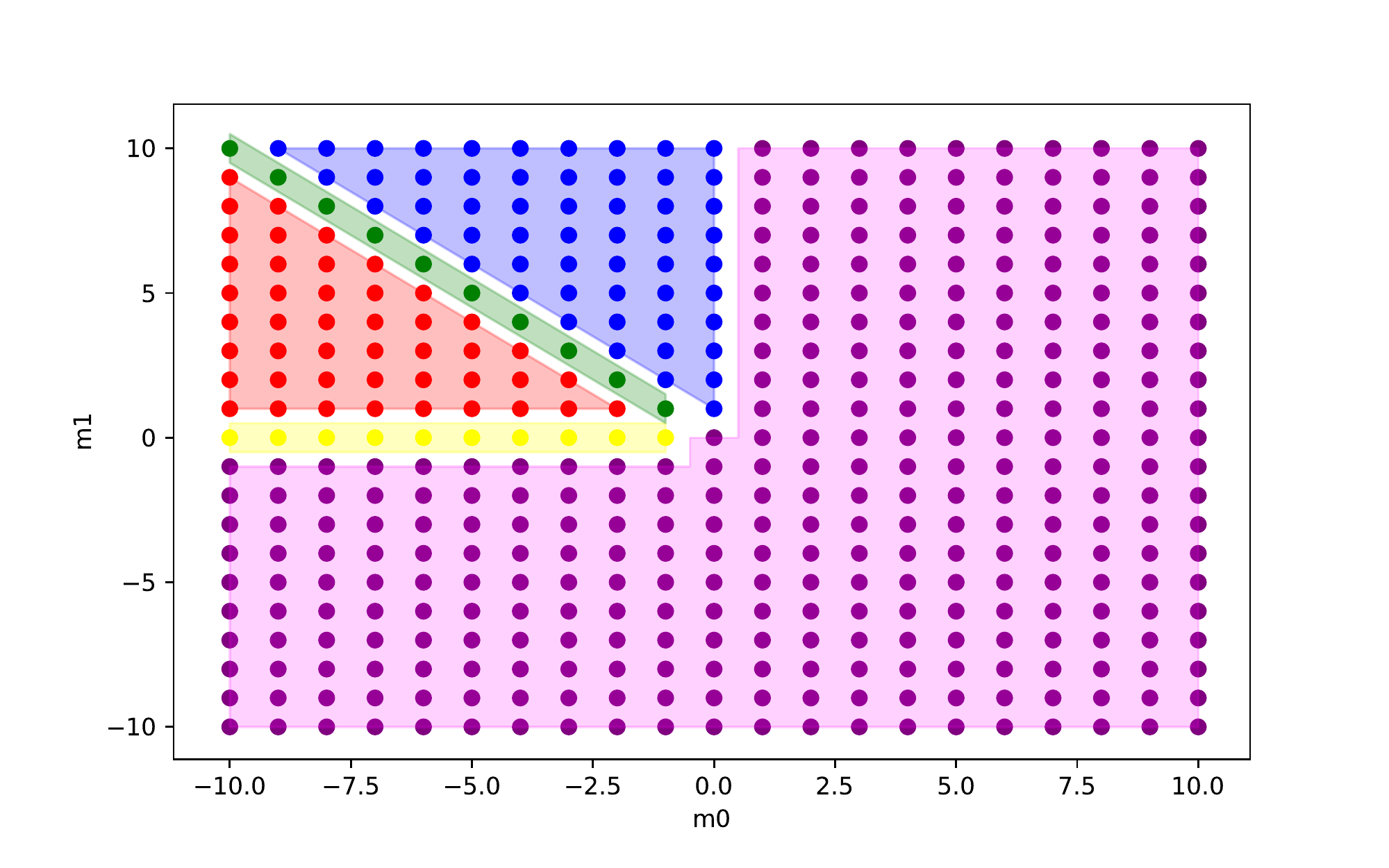} 
		\end{minipage}
		\caption{$h^0$ and $h^1$ for line bundles with charges $m_0$ and $m_1$ over the CICY \# 7888. Colors indicate regions with different polynomial description found in \eqref{eq: 7888h0} and \eqref{eq: 7888h1}.  Near the $m_1$ axis, there are several polynomial descriptions for $h^0$, which leads to an overlap between the blue and the red cone, and the green and the yellow cone.} 
		\label{plt: 7888b}
		\end{figure}
	
	Consider the manifold
	\begin{align}
	M_{7888} =  \left[
	\begin{array}{c||cc}
	1 & 0 & 2  \\
	4 & 4  & 1 
	\end{array}
	\right]^{2,86}_{-168}.
	\end{align}
	Its index is given by
	\begin{align}
	\ind (L) = 2 m_0 m_1^2 + 2m_0 + \frac{4}{3} m_1^3 + \frac{14}{3} m_1
	\end{align}
	and the slope is
	\begin{align}
		\mu (L) = (4m_0 + 8m_1) t_1^2 + 8 m_1 t_0 t_1.
	\end{align}
	Analysing the data yields
	\begin{align}
	\label{eq: 7888h0}
	h^0(X, L) = \begin{cases}
	1+m_0, & m_0 \geq 0, m_1 = 0 \\
	1 - m_0, & m_0 \leq 0,  -m_0 = m_1 \\
	\ind (L) + \frac{2}{3} m_0 (1 - m_0^2), & m_0 \leq 0,  -m_0 < m_1 \\
	\ind(L), & m_0 > 0, m_1 > 0 \\
	0 &\oth 
	\end{cases}
	\end{align}
	and
	\begin{align}
	\label{eq: 7888h1}
	h^1(X, L) = \begin{cases}
	-1-m_0, & m_0 < 0, m_1 = 0 \\
	- \ind (L) +1 -m_0, & m_0 < 0,  -m_0 = m_1 \\
	 \frac{2}{3} m_0 (1 - m_0^2 ), & m_0 \leq 0,  -m_0 < m_1 \\
	- \ind(L), & m_0 < 0,  0 < m_1 < -m_0  \\
	0 &\oth .
	\end{cases}
	\end{align}
	
	Again, there is a clear similarity between both the polynomial expressions and the partition of the line bundle charge space between this example and the previous two. This is particularly striking from figure \ref{plt: 7888b}, where we present the result of this subsection graphically.

	\newpage
	\subsection{CICY manifolds with $[2||2 1]$}
	 \label{sec:221}
	 We now turn to CICY manifolds that have configuration matrices that contain $[2||2 1]$, whose cohomology dimensions are found to  not always admit polynomial descriptions.

	\subsubsection{7833}
	
			\begin{figure}
		\centering
		\begin{minipage}{0.5\textwidth}
			\centering
			\includegraphics[width=1.1\textwidth]{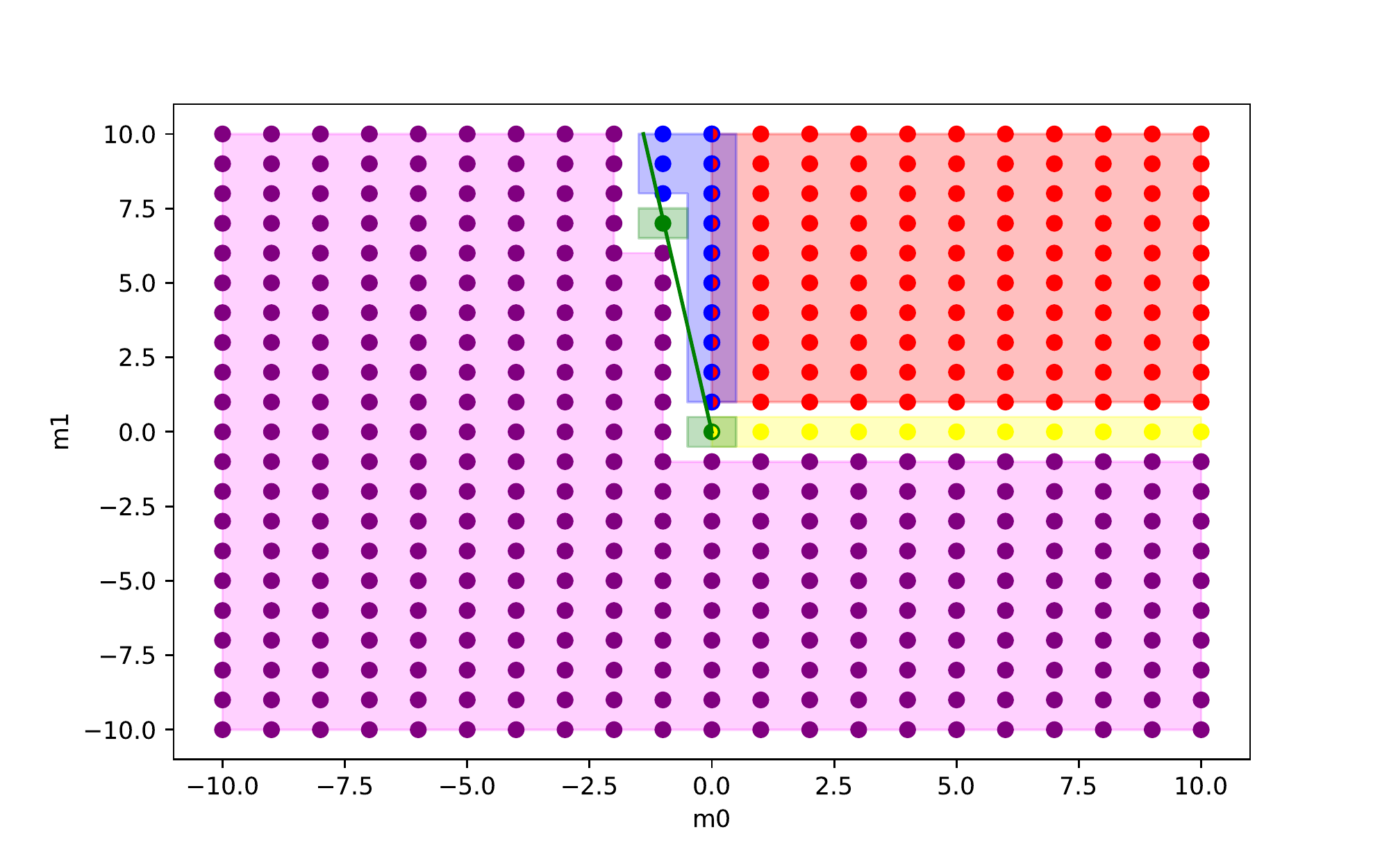} 
		\end{minipage}\hfill
		\begin{minipage}{0.5\textwidth}
			\centering
			\includegraphics[width=1.1\textwidth]{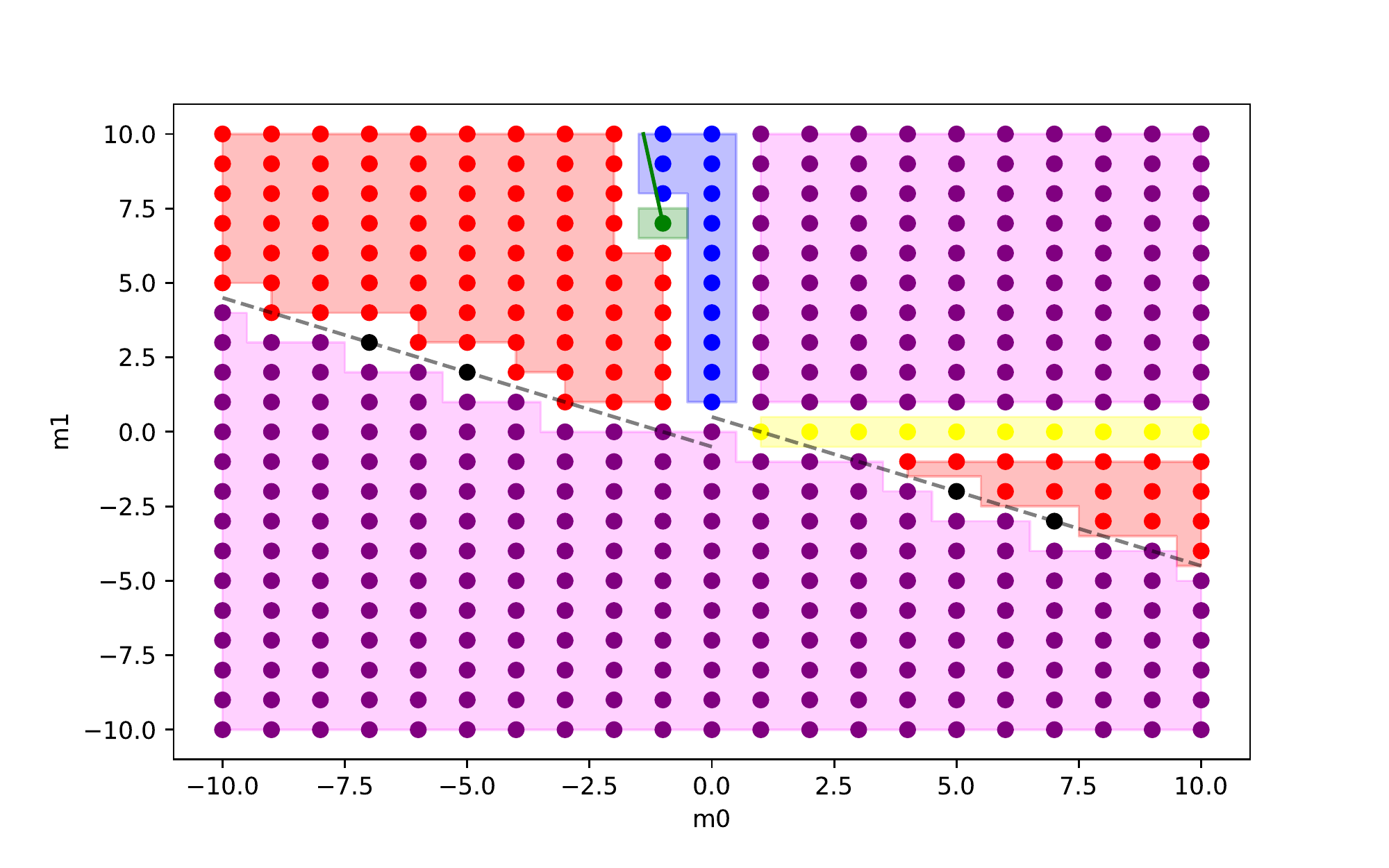} 
		\end{minipage}
		\caption{$h^0$ and $h^1$ for line bundles with charges $m_0$ and $m_1$ over the CICY \# 7833. Colours indicate the different regions and their polynomial description found in \eqref{eq: 7833h0} and \eqref{eq: 7833h1}. The black points on the lines $m_0 = -2m_1-1,
		\; m_0 = -2m_1+1$ are special: the cohomology dimensions in these points disagree with the polynomial expressions of the adjacent regions. }
		\label{plt: 7833b}
	\end{figure}
	
	Consider the manifold
	\begin{align}
	M_{7833} =  \left[
	\begin{array}{c||cc}
	2 & 2 & 1  \\
	3 & 1  & 3 
	\end{array}
	\right]^{2,59}_{-114}.
	\end{align}
	Its index $\ind (L)$ and slope $\mu(L)$ are given by
	\begin{align}
	\label{in: 7833}
	\ind (L) &= \frac{3}{2}m_0^2 m_1+ \frac{7}{2} m_0 m_1^2 + 3 m_0+ \frac{1}{3} m_1^3 + \frac{11}{3} m_1 \; , \\
		\mu (L) &= (6 m_0 + 14 m_1) t_0 t_1 + (7m_0 +2 m_1) t_1^2 + 3m_1 t_0^2 \; .
	\end{align}
	Analysing the data yields
	\begin{align}
	\label{eq: 7833h0}
	h^0(X, L) = \begin{cases}
	\frac{1}{2} (1 + m_0)(2+m_0), & m_0 \geq 0, m_1 = 0 \\
	\frac{1}{2} (1 - m_0)(2-m_0), & m_0 \leq 0, m_1 = -7m_0 \\
	\ind(L) + \frac{1}{3} m_0 (59 - 140 m_0^2), & m_0 \leq 0, m_1 > -7m_0 \\
	\ind(L), & m_0 > 0, m_1 > 0 \\
	0 &\oth 
	\end{cases}
	\end{align}
	and
	\begin{align}
	\label{eq: 7833h1}
	h^1(X, L) = \begin{cases} 
	\frac{1}{2} (1 - m_0)(2-m_0), & m_0 > 0, m_1 = 0 \\
	-\ind(L)  
	+ \frac{1}{2} (1 - m_0)(2-m_0) , & m_0 < 0, m_1 = -7m_0 \\
	 \frac{1}{3} m_0 (59 - 140 m_0^2) , & m_0 \leq 0, m_1 > -7m_0 \\
	-\ind(L), & \begin{cases}
	m_0 < 0, \;  0 < -\frac{1}{2} m_0 -1 < m_1 < -7m_0 \\
	m_0 > 3,  \; 0 > m_1 > -\frac{1}{2}m_0+\frac{1}{2} 
	\end{cases} \\
	0 &\oth .
	\end{cases}
	\end{align}
	Let us make some observations of this result, which is presented graphically in figure \ref{plt: 7833b}. As dictated by the  Kodaira vanishing theorem, the only non-vanishing cohomology dimension in the interior of the first quadrant is $h^0 = \ind(L)$. On the boundary between the first and the fourth quadrant ($m_0\geq 0, m_1=0$), both $h^0$ and $h^1$ are non-zero. In the second quadrant, there is a region between the lines $m_1 = 0$ and $m_1=-7m_0$ where both $h^0$ and $h^1$ are non-vanishing, and are described by polynomials of degree three. Beyond this region, for $m_1<-7m_0$, only $h^1$ is non-vanishing, and is given by the index.  Thus far, the line bundle cohomology follows the pattern observed in the previous section. 
	
	However, there are clear differences with respect to the examples studied in section \ref{sec:102}. First, there is now also a region in the fourth quadrant where $h^1$ is given by the index. Here, we have that the map in \eqref{eq: 234h1} is no longer injective. After passing the line $m_0 = -2m_1+1$, however, the kernel becomes trivial again. On the other side in the second quadrant this implies, by Serre duality, that we have perfect cancellation in kernel and image of the maps in \eqref{eq: 232h1}. 
	
	More puzzling is that the $h^1$ plot has two special points
	\begin{align}
		h^\bullet (X, \mathcal{O}(-7,3)) = (0,3,2,0) \\
		h^\bullet (X, \mathcal{O}(-5,2)) = (0,3,3,0) 
	\end{align}
	where both $h^1 \neq 0 \neq h^2$. Here, we have no cancellation in kernel and image of the maps in \eqref{eq: 232h1}. 
	These points lie on the two lines, which are Serre dual,
	\begin{align}
		m_0 = -2m_1-1 \\
		m_0 = -2m_1+1
	\end{align}
	but there is no common polynomial description for the cohomology dimensions on these lines. Instead, for all points on these lines, apart from the two special ones, the  dimensions are either given by $0$ or $-\ind(L)$. We have checked that this still holds for larger line bundle charges.\footnote{We have checked the line bundle charges (11,-5), (13,-6), (15,-7) and their Serre duals, which are all described by $0$ or $-\ind(L)$.} We note that the Leray tableaux changes significantly between these two black points and their neighbours, and will come back to discuss these irregular points in section \ref{sec:4}. 	
	By Serre duality, this behaviour is repeated for $h^2$ and $h^3$, but with a reflection about the origin. 

\newpage
	\subsubsection{7844}
	
		\begin{figure}
		\centering
		\begin{minipage}{0.5\textwidth}
			\centering
			\includegraphics[width=1.1\textwidth]{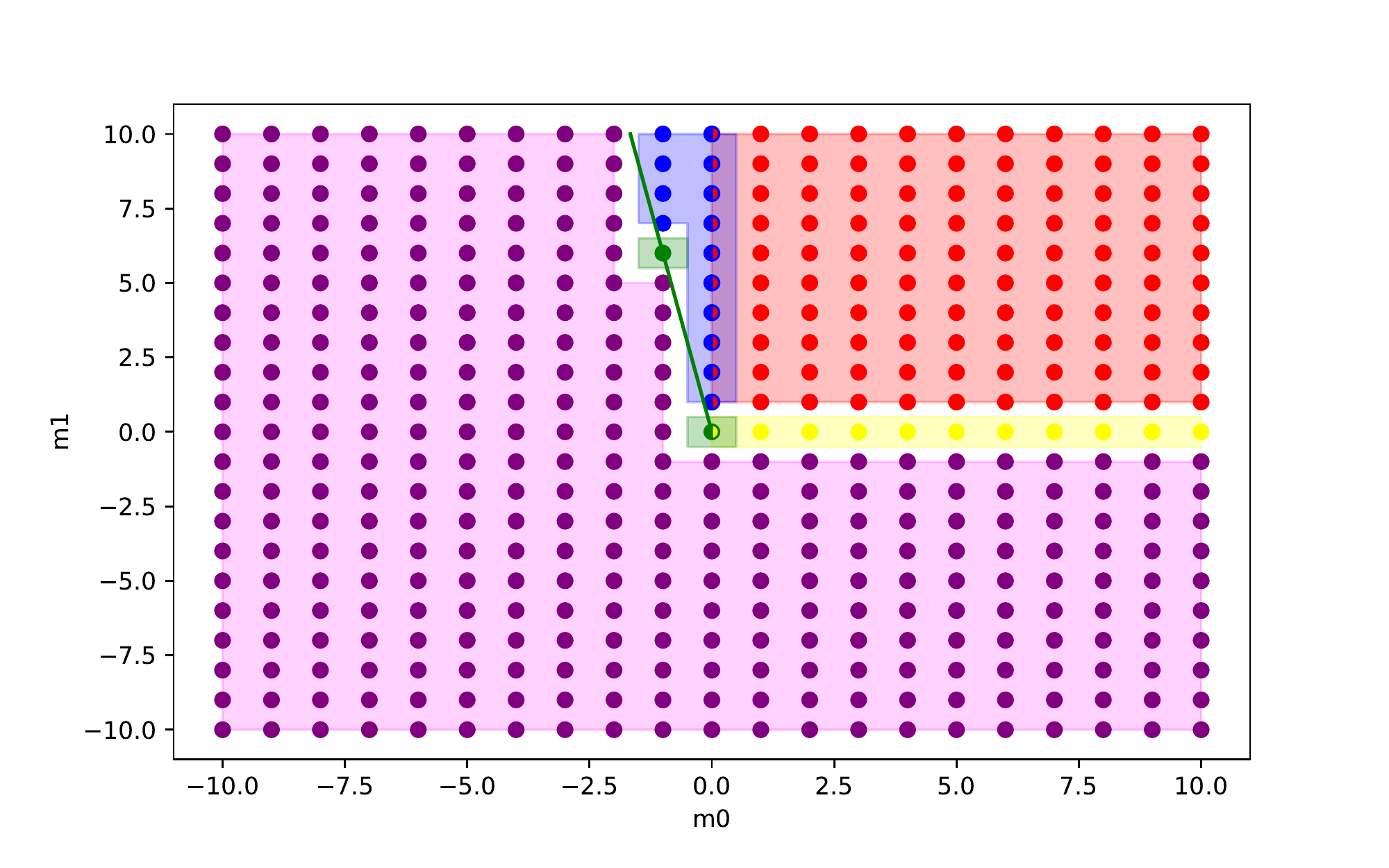} 
		\end{minipage}\hfill
		\begin{minipage}{0.5\textwidth}
			\centering
			\includegraphics[width=1.1\textwidth]{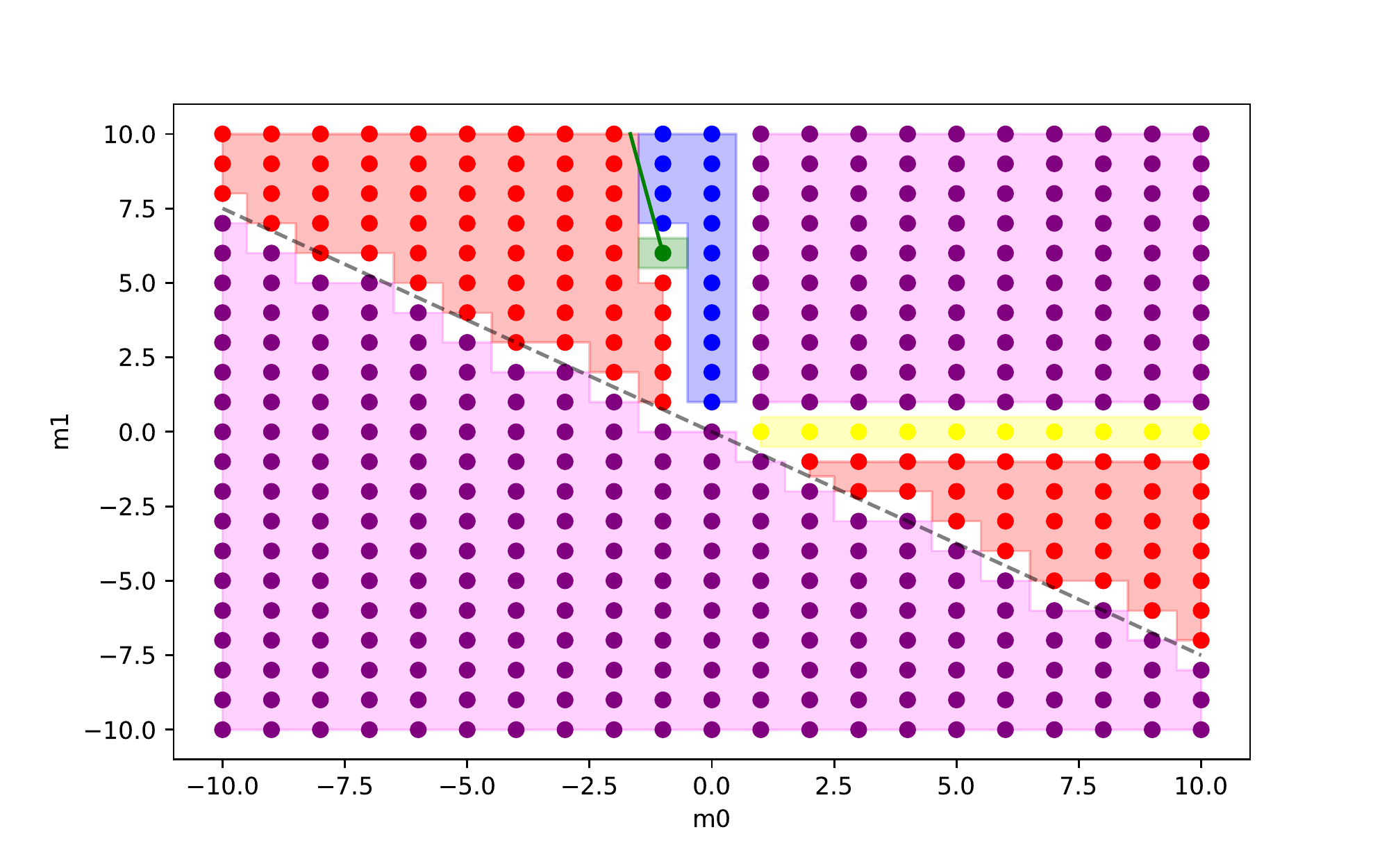} 
		\end{minipage}
		\caption{$h^0$ and $h^1$ for line bundles with charges $m_0$ and $m_1$ over the CICY \# 7844. Colors indicate  different regions with polynomial description found in \eqref{eq: 7844h0} and \eqref{eq: 7844h1}.} 
		\label{plt: 7844b}
	\end{figure}
	
	Consider the manifold
	\begin{align}
	M_{7844} =  \left[
	\begin{array}{c||cc}
	2 & 2 & 1  \\
	3 & 2  & 2 
	\end{array}
	\right]^{2,62}_{-120}.
	\end{align}
	The index is given by
	\begin{align}
	\label{in: 7844}
	\ind(L) = 2m_0^2 m_1 + 3 m_0 m_1^2 + 3 m_0 + \frac{1}{3} m_1^3 + \frac{11}{3} m_1
	\end{align}
	and the slope is
	\begin{align}
		\mu (L) = (8 m_0 + 12 m_1) t_0 t_1 + (6m_0 + 2m_1) t_1^2 + 4m_1 t_0^2 .
	\end{align}
	Analysing the data yields
	\begin{align}
	\label{eq: 7844h0}
	h^0(X, L) = \begin{cases}
	\frac{1}{2} (1 + m_0)(2+m_0), & m_0 \geq 0, m_1 = 0 \\
	\ind(L), & m_0 \geq 0, m_1 > 0 \\
	\frac{1}{2}(1-m_0)(2-m_0), 
	& m_0 < 0, m_1 = -6 m_0 \\
	8m_0(2-3m_0^2) + \ind (L), & m_0 < 0, m_1 > -6 m_0 \\
	0 &\oth 
	\end{cases}
	\end{align}
	and
	\begin{align}
	\label{eq: 7844h1}
	h^1(X, L) = \begin{cases}
	\frac{1}{2} (1 - m_0)(2-m_0), & m_0 > 0, m_1 = 0 \\
	-\ind(L) & \begin{cases}
	m_0 \geq 3, - \frac{3}{4} m_0 < m_1 < 0  \\
	m_0 < 0, -6m_0 > m_1 \geq - \frac{3}{4} m_0  > 0 \\
	\end{cases}\\
	-\ind(L)+ \frac{1}{2}(1-m_0)(2-m_0),
	& m_0 < 0, m_1 = -6 m_0 \\
	8m_0(2-3m_0^2), & m_0 < 0, m_1 > -6 m_0 \\
	0 &\oth \; .
	\end{cases}
	\end{align}
	This CICY manifold was studied in \cite{Constantin:2018hvl}, where polynomial expressions for the line bundle cohomology dimensions were reported. Our investigations confirm these results as shown in figure \ref{plt: 7844b}.\footnote{There are two minor differences with respect to \cite{Constantin:2018hvl}. First, we have a  green boundary in the second quadrant, at the line $ m_1 = -6 m_0$. This was not seen in Ref \cite{Constantin:2018hvl}, since for the points (-1,6) and (-2,12) the blue and green polynomials coincide, but for larger charges we observe a discrepancy. Second, the boundary between the red and purple region, where the dimension is given by either 0 or the index, is shifted as compared to Ref \cite{Constantin:2018hvl}, which again is only noticeable at larger charges. We thank the authors of Ref \cite{Constantin:2018hvl} for valuable discussions on these points.}
	Note that all points can be described by polynomials in this example; there are no black points on the boundaries between regions.

	\subsubsection{7883}
	
	\begin{figure}
		\centering
		\begin{minipage}{0.5\textwidth}
			\centering
			\includegraphics[width=1.1\textwidth]{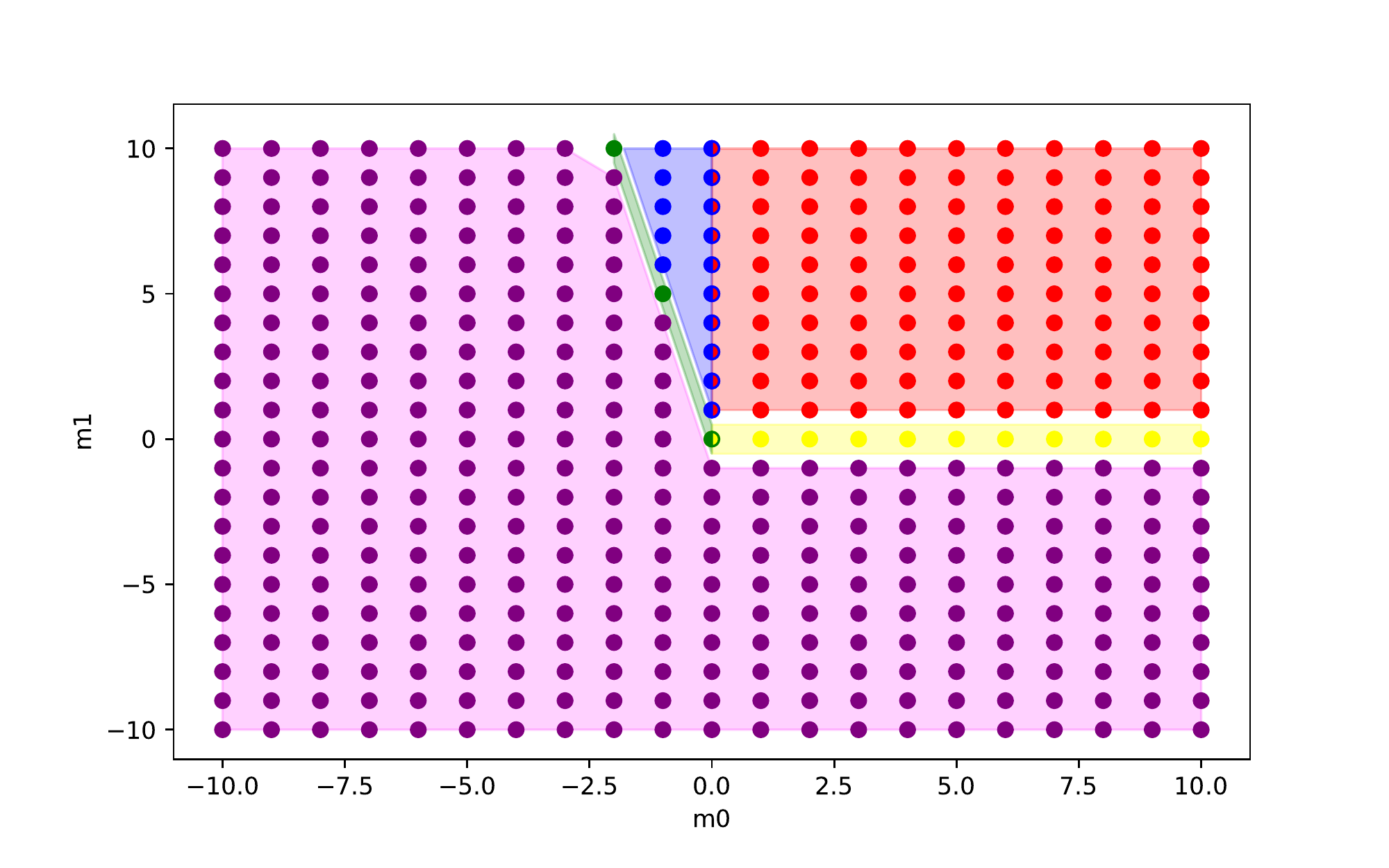} 
		\end{minipage}\hfill
		\begin{minipage}{0.5\textwidth}
			\centering
			\includegraphics[width=1.1\textwidth]{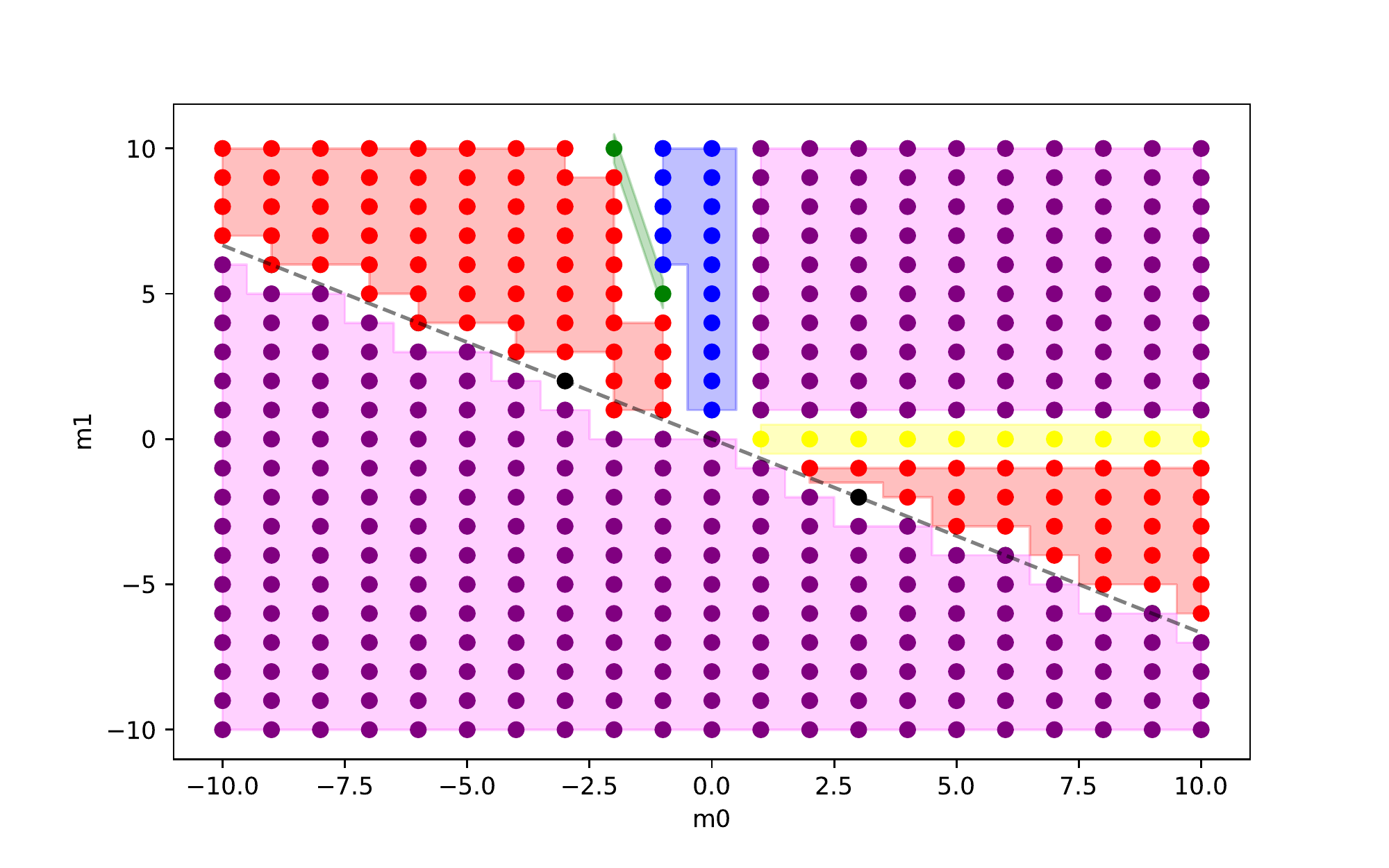} 
		\end{minipage}
		\caption{$h^0$ and $h^1$ for line bundles with charges $m_0$ and $m_1$ over the CICY \# 7883. Colors indicate the different regions and their polynomial description found in \eqref{eq: 7883h0} and \eqref{eq: 7883h1}. Black points again indicate that a polynomial fit is lacking.} 
		\label{plt: 7883b}
	\end{figure}
	
	Consider the manifold
	\begin{align}
	M_{7833} =  \left[
	\begin{array}{c||cc}
	2 & 2 & 1  \\
	3 & 3  & 1 
	\end{array}
	\right]^{2,77}_{-150}.
	\end{align}
	Its index is given by
	\begin{align}
	\label{in: 7883}
	\ind (L) = \frac{3}{2} m_0^2 m_1 +\frac{5}{2} m_0 m_1^2 + \frac{1}{3} m_1^3 + \frac{11}{3} m_1
	\end{align}
	and the slope is
	\begin{align}
		\mu (L) = (6m_0+10m_1) t_0 t_1 + (5m_0 + 2 m_1) t_1^2 + 3m_1 t_0^2. 
	\end{align}
	Analysing the data yields
	\begin{align}
	\label{eq: 7883h0}
	h^0(X, L) = \begin{cases}
	\frac{1}{2} (1 + m_0)(2+m_0), 
	& m_0 \geq 0, m_1 = 0 \\
	\frac{1}{2} (1 - m_0)(2-m_0),  
	& m_0 \leq 0, m_1 = -5m_0 \\
	\frac{1}{3}m_0(37-40m_0^2)+\ind(L), & m_0 \leq 0, m_1 > -5m_0 \\
	\ind(L), & m_0 \geq 0, m_1 > 0 \\
	0 &\oth
	\end{cases}
	\end{align}
	and
	\begin{align}
	\label{eq: 7883h1}
	h^1(X, L) = \begin{cases}
	\frac{1}{2} (1 - m_0)(2-m_0), 
	& m_0 > 0, m_1 = 0 \\
	-\ind(L)+\frac{1}{2} (1 - m_0)(2-m_0), 
	& m_0 < 0, m_1 = -5m_0 \\
	\frac{1}{3}m_0(37-40m_0^2),
	 & m_0 < 0, m_1 > -5m_0 \\
	-\ind(L), & \begin{cases}
	m_0 < 0, - \frac{2}{3} m_0 \leq m_1 < -5m_0 \\
	m_0 > 2,  0 < - m_1 < \frac{2}{3} m_0 \\
	\end{cases}\\
	0 &\oth 
	\end{cases}
	\end{align}
	In figure \ref{plt: 7883b} we plot this result. We note that there are again a couple of problematic points. First
	the point (-2,1) is described by the index but lies beneath the separating line of
	\begin{align}
		m_0 = - \frac{3}{2} m_1.
	\end{align}
	Second, the point
	\begin{align}
		h^\bullet (X, \mathcal{O}(-3,2)) = (0,3,1,0)
	\end{align} 
	belongs neither to the red index region nor to the trivial purple one as both $h^1, h^2$ are nonzero. Checking further points outside the range $\{-10,10\}$ we encounter more black points with non vanishing $h^1, h^2$ sitting on $m_0 = - \frac{3}{2} m_1$. Those are \footnote{Note: The calculation of $h^\bullet (X, \mathcal{O}(-18,12))$ took 16 hours on 20 cores with $>100$ GB memory usage. Due to these computational limitations, we were unable to study points with higher charges on this line.}
	\begin{align}
		&h^\bullet (X, \mathcal{O}(-12,8)) = (0,29,1,0) \\
		&h^\bullet (X, \mathcal{O}(-15,10)) = (0,127,77,0) \\
		&h^\bullet (X, \mathcal{O}(-18,12)) = (0,392,310,0). 
	\end{align}
	We were unable to find a converging polynomial fit through these special points.

	\newpage
	\subsection{CICY manifolds with $[1||11]$}
	\label{sec:111}
	The last set of CICY manifolds  have configuration matrices that contain $[1||1 1]$. Here we find that the cohomology dimensions  display polynomial behaviour in some regions, and a a novel recursive behaviour in some.

	\subsubsection{7858}

	Consider the manifold,
	\begin{align}
	M_{7858} =  \left[
	\begin{array}{c||cc}
	1 & 1 & 1  \\
	4 & 3  & 2 
	\end{array}
	\right]^{2,66}_{-128}.
	\end{align}
	Its index and slope are given by
	\begin{align}
	\ind (L) = 3 m_0 m_1^2 +2 m_0+ \frac{5}{6} m_1^3 + \frac{25}{6} m_1 \; , \\
		\mu (L) = (6m_0 + 5 m_1) t_1^2 + 12 m_1 t_0 t_1 \; .
	\end{align}
	Analysing the data yields
	\begin{align}
	\label{eq: 7858h0}
	h^0(X, L) = \begin{cases}
	1+m_0, & m_0 \geq 0, m_1 = 0 \\
	1, & m_0 \leq 0, m_1 = -2m_0 \\
	\frac{3}{2}(2m_0 +m_1)\left(3+(2m_0 +m_1)^2 \right) , & m_0 < 0,  -2m_0 < m_1 < -3m_0 \\
	6m_0(1-m_0^2)+\ind(L), & m_0 \leq 0, -3m_0 \leq m_1  \\
	\ind(L), & m_0 > 0, m_1 > 0 \\
	0 &\oth 
	\end{cases}
	\end{align}
	and
	\begin{align}
	\label{eq: 7858h1}
	h^1(X, L) = \begin{cases}
	-1-m_0, & m_0 < 0, m_1 = 0 \\
	-\ind(L)+1, & m_0 < 0, m_1 = -2m_0 \\
	-\ind(L) + \frac{3}{2}(2m_0 +m_1)\left(3+(2m_0 +m_1)^2 \right) ,
	& m_0 < 0,  -2m_0 < m_1 < -3m_0 \\
	6m_0(1-m_0^2), & m_0 \leq 0, -3m_0 \leq m_1  \\
	-\ind(L), & m_0 < 0, 0 < m_1 < -2m_0 \\
	0 &\oth \; .
	\end{cases}
	\end{align}
	
			\begin{figure}
			\centering
			\begin{minipage}{0.5\textwidth}
				\centering
				\includegraphics[width=1.1\textwidth]{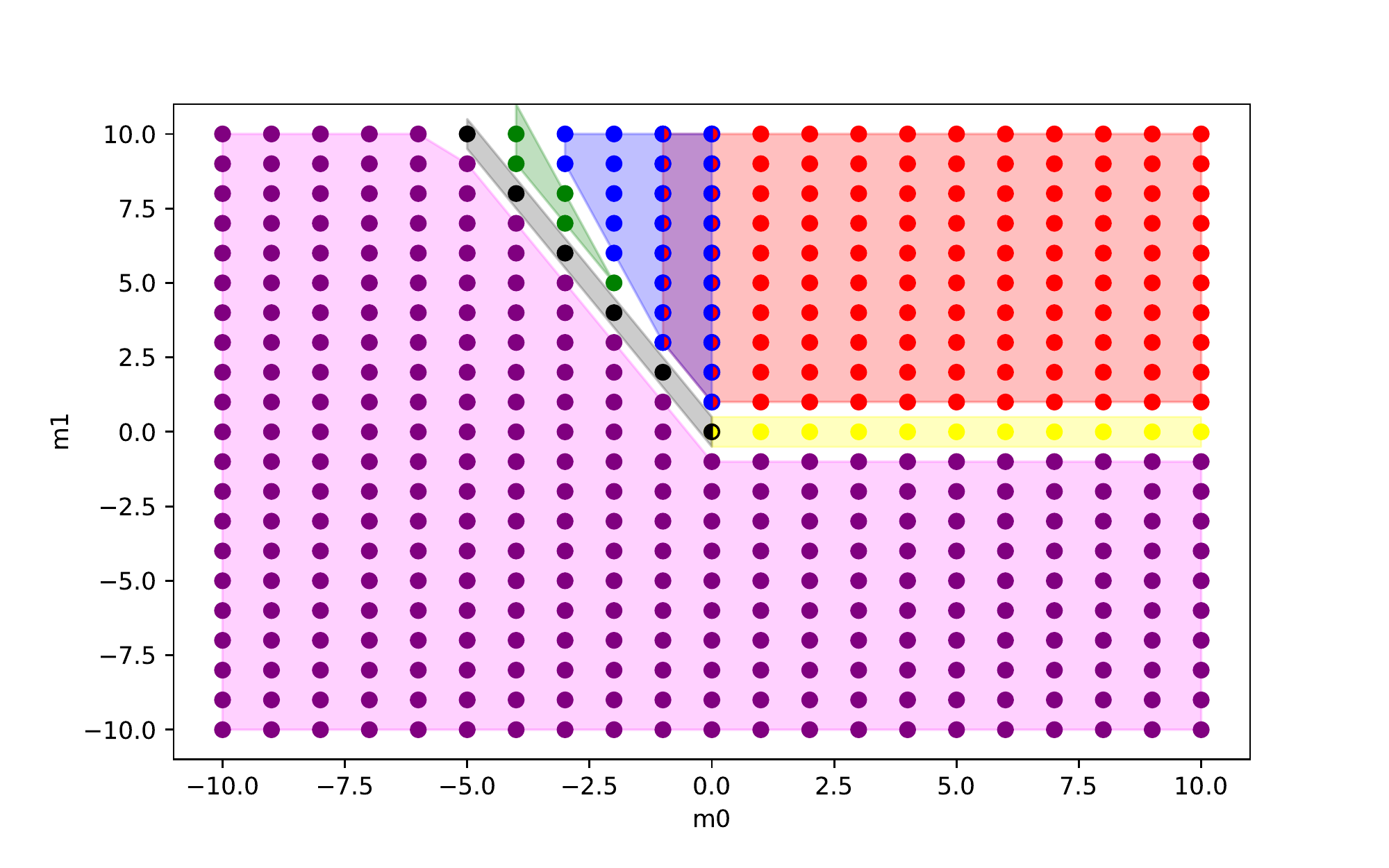} 
			\end{minipage}\hfill
			\begin{minipage}{0.5\textwidth}
				\centering
				\includegraphics[width=1.1\textwidth]{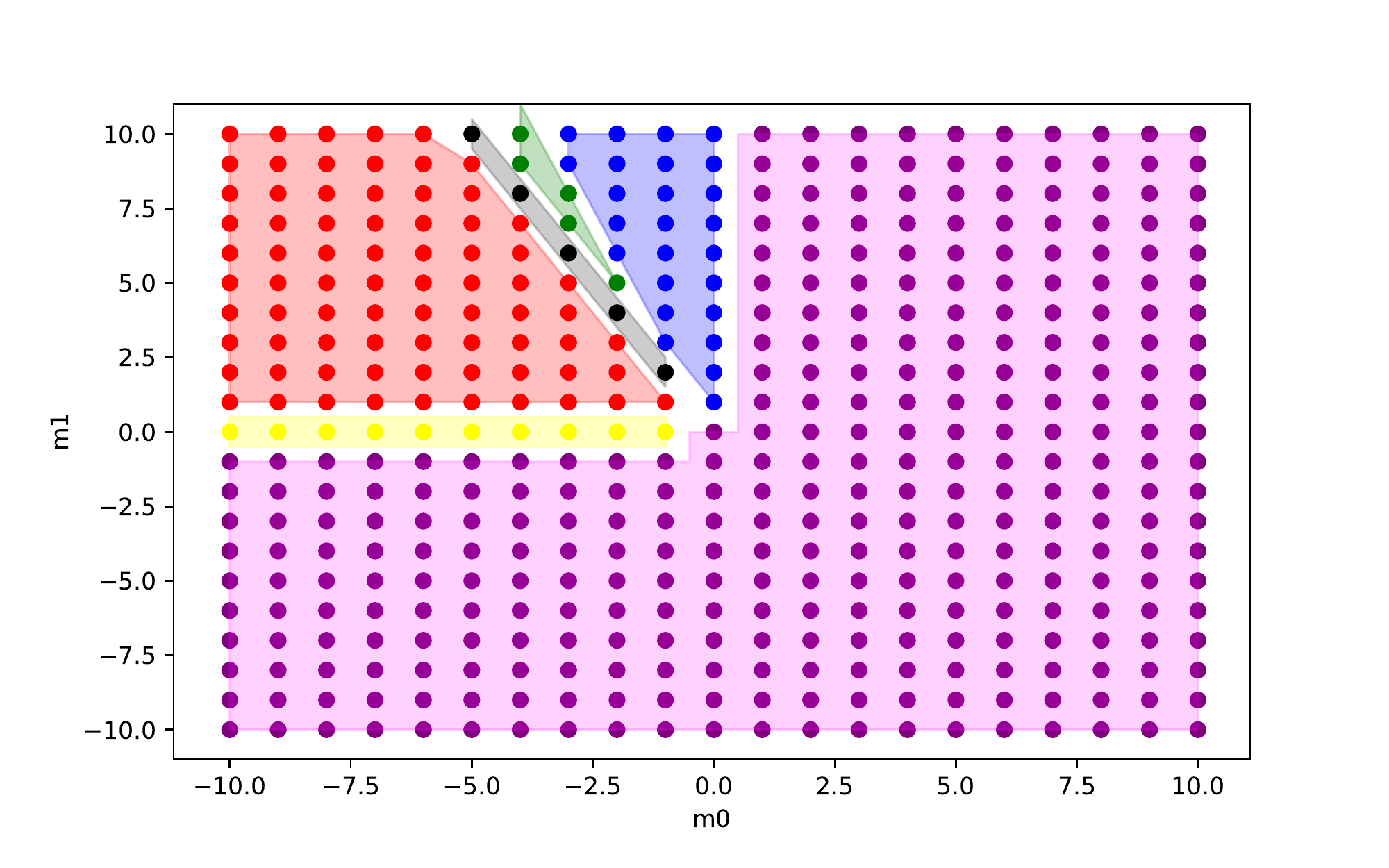} 
			\end{minipage}
			\caption{$h^0$ and $h^1$ for line bundles with charges $m_0$ and $m_1$ over the CICY \# 7858. Colors indicate the different regions with polynomial description found in \eqref{eq: 7858h0} and \eqref{eq: 7858h1}.}
			\label{plt: 7858} 
		\end{figure}
		
		\begin{figure}
			\centering
			\begin{minipage}{0.5\textwidth}
				\centering
				\includegraphics[width=1.1\textwidth]{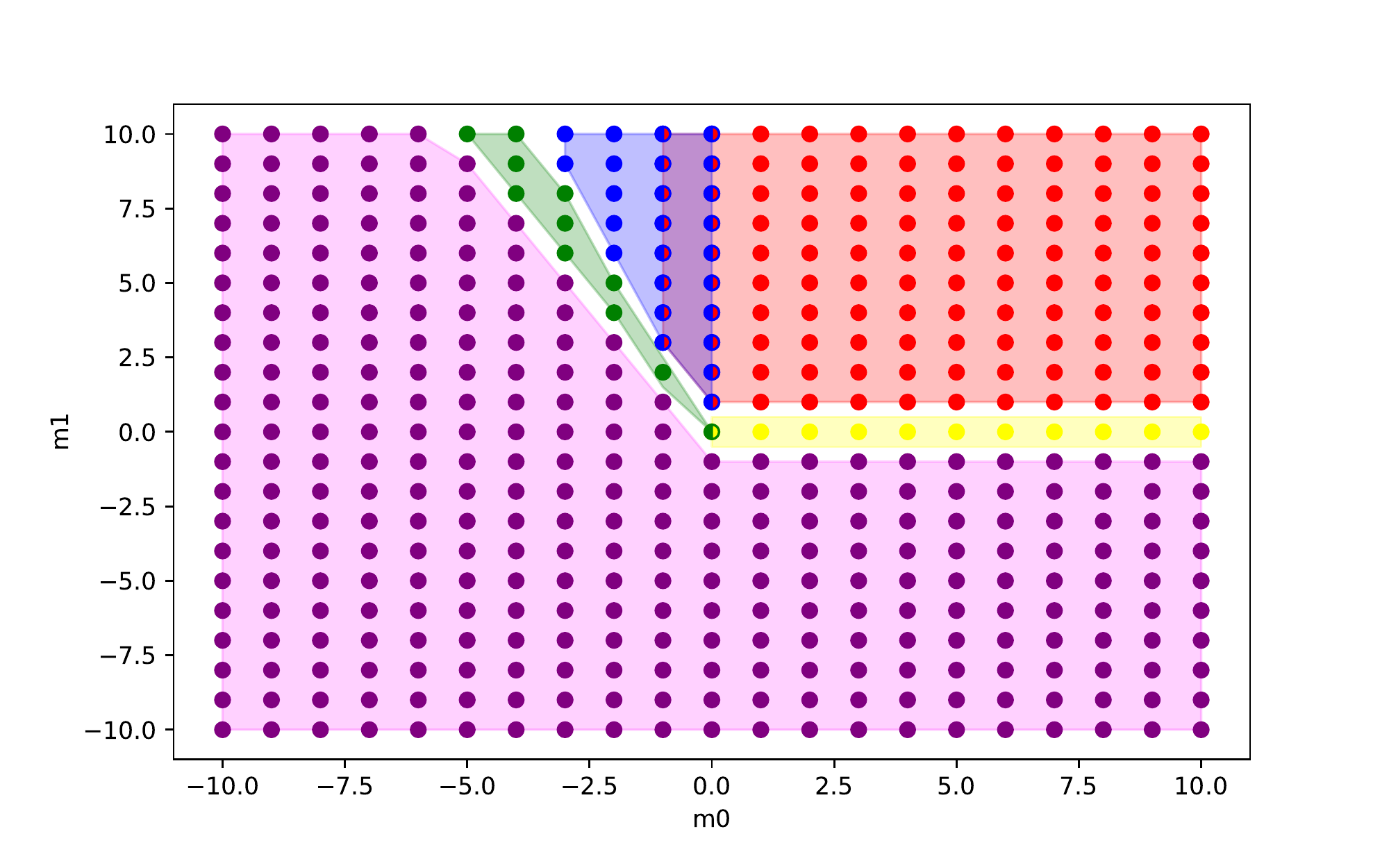} 
			\end{minipage}\hfill
			\begin{minipage}{0.5\textwidth}
				\centering
				\includegraphics[width=1.1\textwidth]{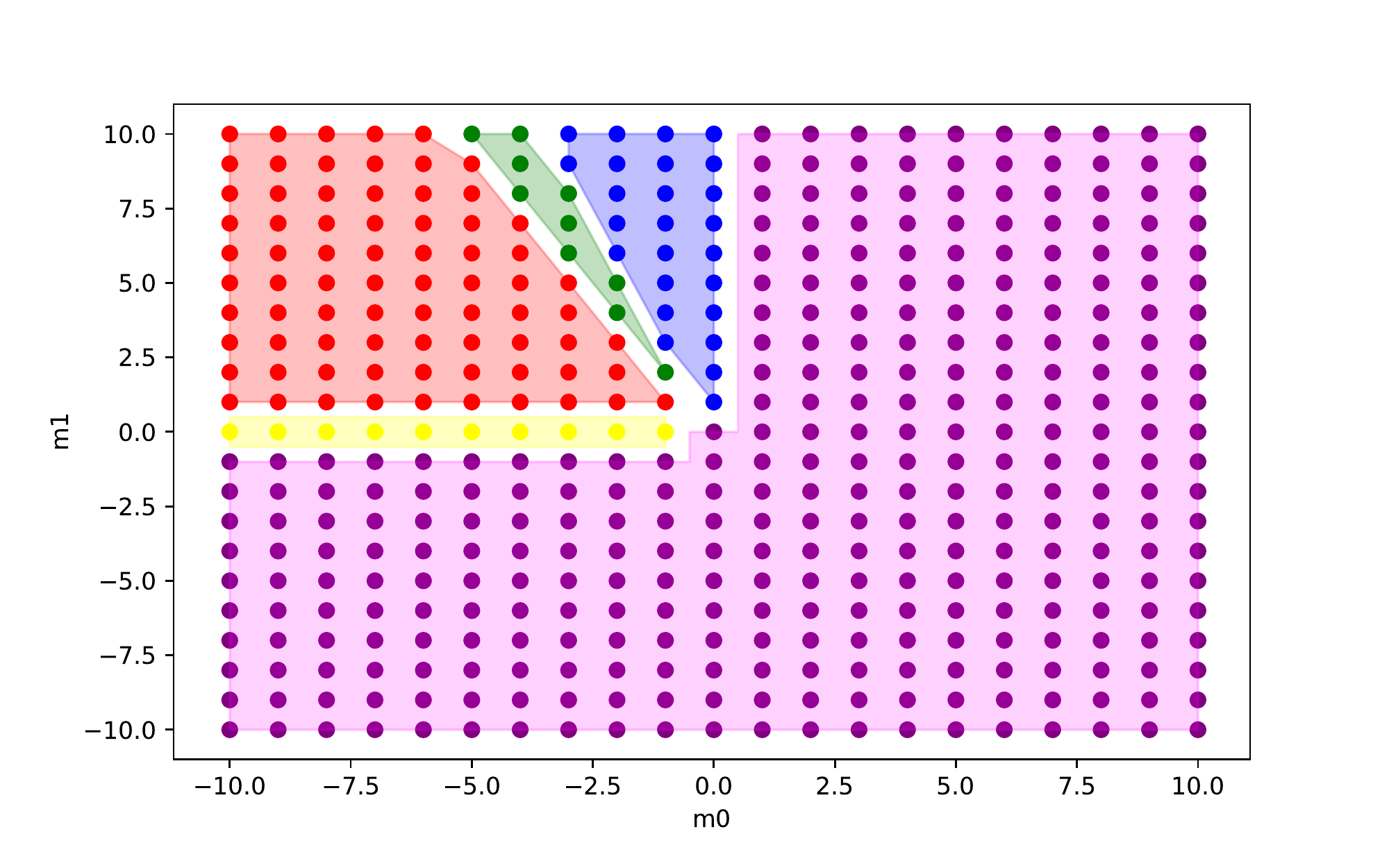} 
			\end{minipage}
			\caption{$h^0$ and $h^1$ for line bundles with charges $m_0$ and $m_1$ over the CICY \# 7858. Colors indicate the different regions and their polynomial or recursive description found in \eqref{eq: 7858h0r} and \eqref{eq: 7858h1r}. }
			\label{plt: 7858b}
		\end{figure}
	
	The result is presented in figure \ref{plt: 7858}. A more elegant description of the cohomology dimensions is given recursively by
	\begin{align}
	\label{eq: 7858h0r}
	h^0(X, L) = \begin{cases}
	1+m_0, & m_0 \geq 0, m_1 = 0 \\
	h^0(X, L(m_0+1, m_1-2)), & m_0 < 0, -3m_0 > m_1 \geq -2m_0 \\
	6m_0(1-m_0^2)+\ind(L), & m_0 \leq 0, -3m_0 \leq m_1  \\
	\ind(L), & m_0 > 0, m_1 > 0 \\
	0 &\oth 
	\end{cases}
	\end{align}
	and
	\begin{align}
	\label{eq: 7858h1r}
	h^1(X, L) = \begin{cases}
	-1-m_0, & m_0 < 0, m_1 = 0 \\
	h^0(X, L(m_0+1, m_1-2))-\ind(L), & m_0 < 0, -3m_0 > m_1 \geq -2m_0 \\
	6m_0(1-m_0^2), & m_0 \leq 0, -3m_0 \leq m_1  \\
	-\ind(L), & m_0 < 0, 0 < m_1 < -2m_0 \\
	0 &\oth .
	\end{cases}
	\end{align}
	Graphically the new regions are presented in figure \ref{plt: 7858b}. Due to the recursive relationship, the green region consists of parallel lines of the form
	\begin{align}
		m_1 = -2 m_0 + c, \qquad  c \in \mathbb{Z}_{>0}
	\end{align}
	with constant $h^0$. For this specific manifold we can also find a polynomial description matching the constant values of these lines, excluding the first at $c=0$ and $m_1 = -2 m_0$. This one is expected to have $h^0 = 1$, as the boundary towards the purple region starts at the origin.

	\subsubsection{7885}
	
	\begin{figure}
		\centering
		\begin{minipage}{0.5\textwidth}
			\centering
			\includegraphics[width=1.1\textwidth]{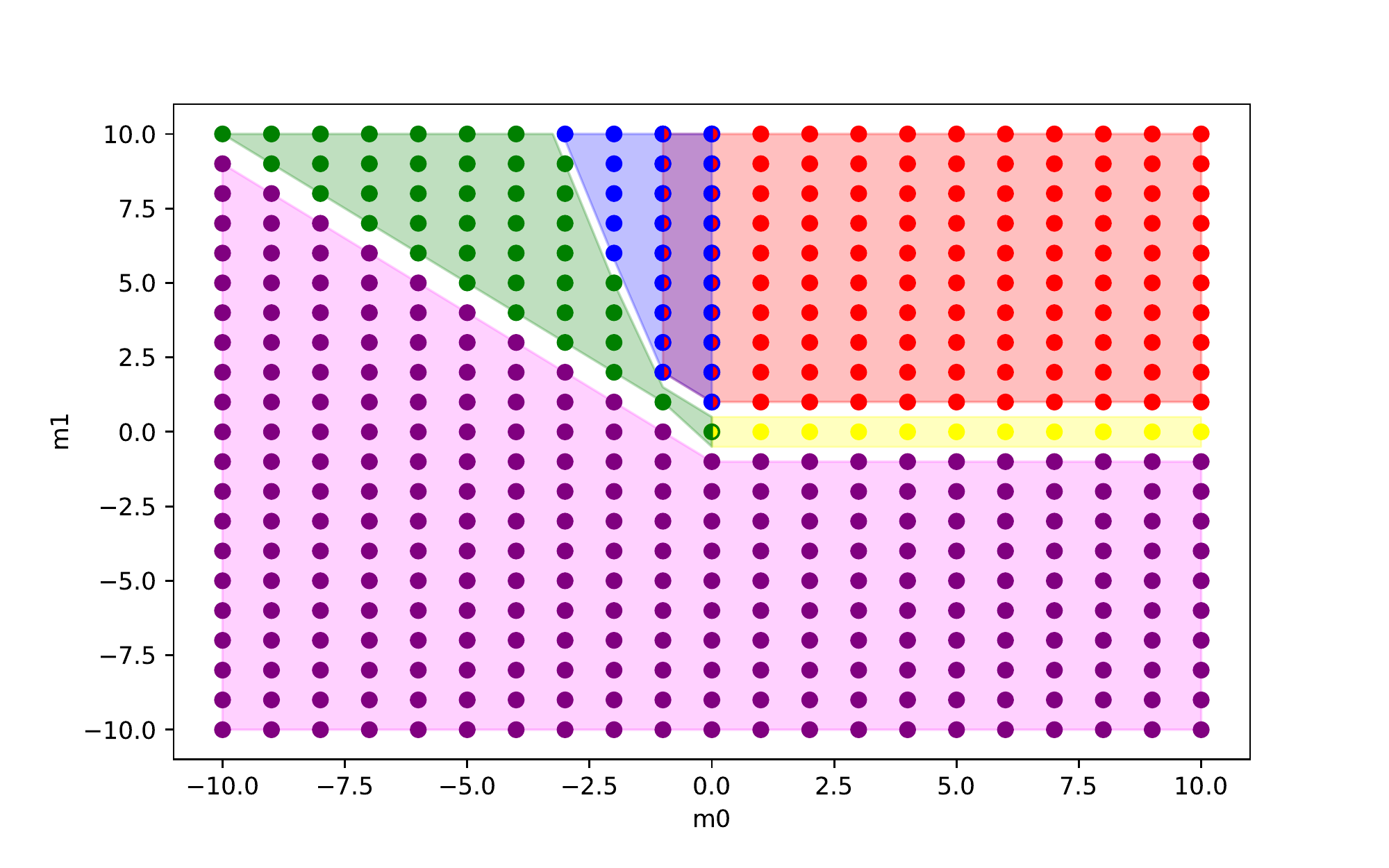} 
		\end{minipage}\hfill
		\begin{minipage}{0.5\textwidth}
			\centering
			\includegraphics[width=1.1\textwidth]{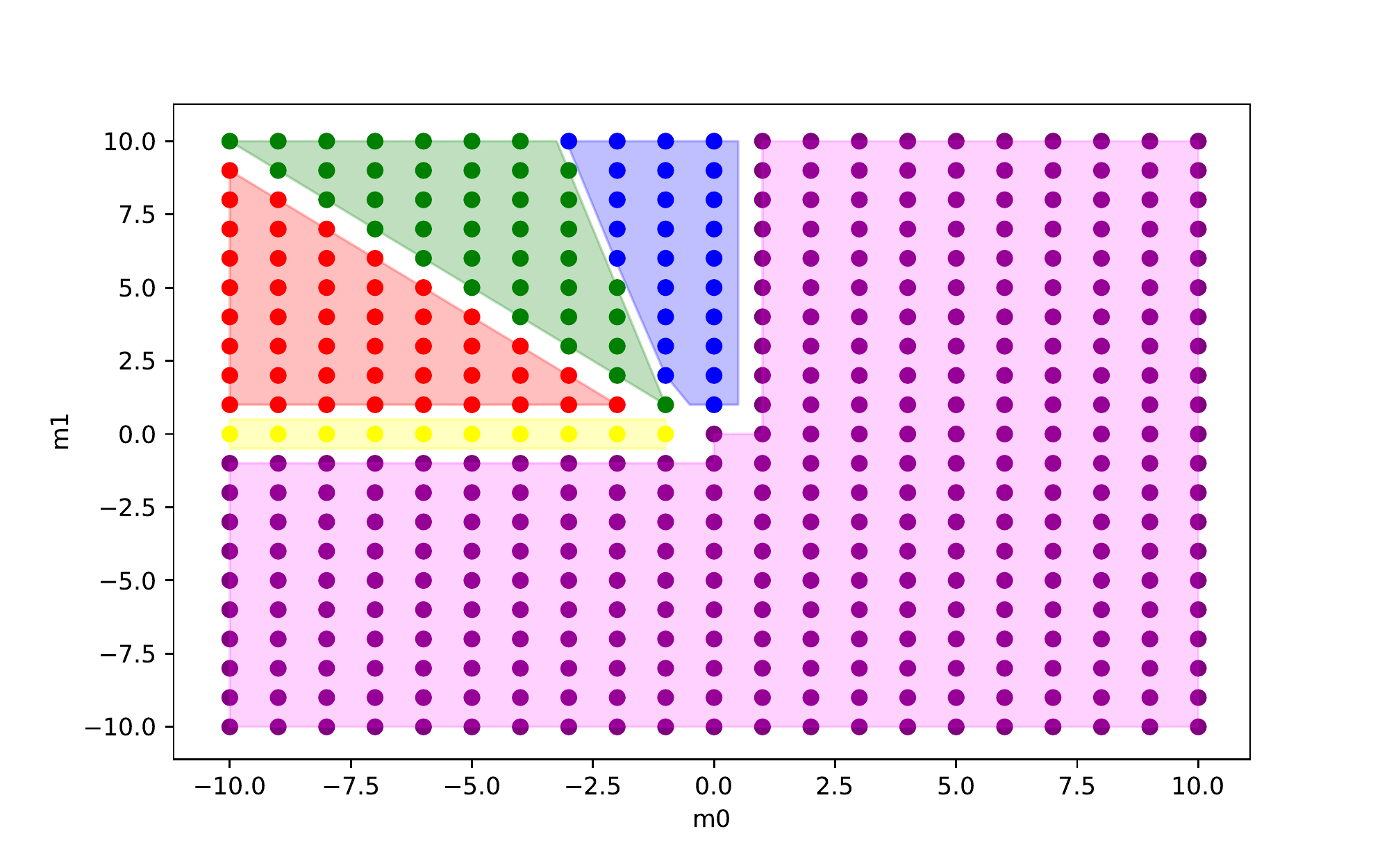} 
		\end{minipage}
		\caption{$h^0$ and $h^1$ for line bundles with charges $m_0$ and $m_1$ over the CICY \# 7885. Colors indicate the different regions and their polynomial description found in \eqref{eq: 7885h0r} and \eqref{eq: 7885h1r}, with some overlaps close to the $m_1$ axis.  For the green region we only have a recursive relationship.} 
		\label{plt: 7885b}
	\end{figure}
	
	Consider the manifold
	\begin{align}
	M_{7885} =  \left[
	\begin{array}{c||cc}
	1 & 1 & 1  \\
	4 & 4  & 1 
	\end{array}
	\right]^{2,86}_{-168}.
	\end{align}
	Its index is given by
	\begin{align}
	\ind (L) = 2 m_0 m_1^2 + 2m_0 + \frac{5}{6} m_1^3 + \frac{25}{6} m_1
	\end{align}
	and the slope is
	\begin{align}
		\mu (L) = (4m_0 + 5m_1) t_1^2 + 8m_1 t_0 t_1 .
	\end{align}
	Analysing the data yields
	\begin{align}
	\label{eq: 7885h0r}
	h^0(X, L) = \begin{cases}
	1+m_0, & m_0 \geq 0, m_1 = 0 \\
	h^0(X,L(m_0+1, m_1-1)), & m_0 < 0,  -4m_0-2  \leq m_1 \\
	\frac{8}{3}m_0(1-m_0^2)+\ind(L), & m_0 \leq 0,  -4m_0-2 \leq m_1, m_1 > 0 \\
	\ind(L), & m_0 > 0, m_1 > 0 \\
	0 &\oth 
	\end{cases}
	\end{align}
	and
	\begin{align}
	\label{eq: 7885h1r}
	h^1(X, L) = \begin{cases}
	-1-m_0, & m_0 < 0, m_1 = 0 \\
	h^0(X,L(m_0+1, m_1-1))-\ind(L), & m_0 < 0, -4m_0-2 > m_1 \geq -m_0 \\
	\frac{8}{3}m_0(1-m_0^2), & m_0 \leq 0,  -4m_0-2 \leq m_1, m1 > 0 \\
	-\ind(L), & m_0 < -1, 0 < m_1 < -m_0 \\
	0 &\oth .
	\end{cases}
	\end{align}
	Here, we were unable to find a converging polynomial fit for the dimensions in the green region. The result is plotted in \ref{plt: 7888b}. Due to the recursive relationships, we again find parallel green lines of the form
	\begin{align}
		m_1 = -m_0 + c, \qquad c \in \mathbb{Z}_{>0}
	\end{align}
	with constant $h^0$. In principle one could now fit for every four neighbouring lines a polynomial, but that would leave us with an infinite number of regions.
	We further emphasise that the boundary between green and blue region is given by the line
	\begin{align}
		m_1 = -4m_0-2
	\end{align}
	which for the first time significantly misses the origin.

	\section{Discussion}
	
	\label{sec:4}
	
	In this section we will analyse our results from the previous section.
	We will split the discussion into two parts. First, we discuss features of the $h^0$-plots and second of the $h^1$ plots.  We recall that we previously defined three sets $G_1 = \{7806, 7882, 7888\}$ sharing the row $[1||02]$ in their configuration matrix, $G_2 = \{7833, 7844, 7883\}$ sharing the row $[2||21]$ and $G_3 = \{ 7858, 7885\}$ sharing the row $[1||11]$.
	
	\subsection{$h^0$ features}

	All $h^0$ plots have several things in common. The first and third quadrant are dictated by Kodaira's vanishing theorem, such that we find a red and purple region,  respectively. Recalling the results from section \ref{sequence}, we found that in the fourth quadrant $h^0 = 0$ for all considered CICYs. This was a simple consequence from the vanishing of ambient space cohomologies after applying the Bott--Borel--Weil theorem.
	
	On the boundary between fourth and first quadrant at $m_1 = 0$ we always find a yellow line, separating purple and red region. It is described by the polynomials
	\begin{align}
		h^0 = \begin{cases}
		1 + m_0 & \text{for } \mathcal{A} = \mathbb{P}^1 \times \mathbb{P}^4 \\
		\frac{1}{2} (m_0 +1) (m_0 +2) & \text{for } \mathcal{A} = \mathbb{P}^2 \times \mathbb{P}^3  
		\end{cases}.
	\end{align}
	On the other hand, when studying the second quadrant, we similarly encounter a boundary, the green region/line. It separates the purple region from the blue one, where $h^0$ becomes non trivial. For the sets $G_1$ and $G_2$ this green line takes the same polynomial description as the yellow one,
	\begin{align}
	h^0 = \begin{cases}
	1 - m_0 & \text{for } G_1 \\
	\frac{1}{2} (m_0 -1) (m_0 -2) & \text{for } G_2
	\end{cases}
	\end{align}
	when accounting for the fact that $m_0$ differs in sign between the green and yellow lines. This is fairly remarkable, as it shows that the trivial purple region is separated from the non zero points by, in principle, a single boundary. Similar boundaries have been found in \cite{Constantin:2018hvl} for almost all investigated manifolds. 
	
	The case of set $G_3$ is slightly different, as the green line here is enhanced to a two-dimensional region, where the cohomology data is described recursively. We also note that the line next to the purple region is always given by $h^0 = 1$, which is expected, since otherwise it can not hit the origin.
	
	Adjacent to the green region, we encounter a blue one where $h^0$ is always described by a polynomial of the form
	\begin{align}
		h^0 = \ind(L) + m_0 (c_1 - c_2m_0^2) \; ,
	\end{align}
	with constant $c_1, c_2$.	Due to this particular form, we find that for all manifolds the index provides an alternative description for the line line $m_0 = 0$ and further for the ambient space $\mathcal{A} = \mathbb{P}^1 \times \mathbb{P}^4$ it also includes the line $m_0 = \pm 1$ as we find here for all examples $c_1 = c_2$. The same form of polynomial is encountered several times in \cite{Constantin:2018hvl}, also for CICYs with higher Picard number. 
	
	We can also make some comments about the position of the green boundary. For $G_{1/3}$ we have by \eqref{eq: 142h0} that the kernel of one map in the Leray tableaux has to cancel exactly with the image of another map. This is rather difficult to study generically without explicitly knowing the involved maps. The best indication for the appearance {of the green boundary} is given by the vanishing theorem described at the end of \ref{sec:2}. We have $h^0 = 0$ whenever the slope becomes zero somewhere in the K\"ahler cone.  For e.g. the manifold 7806, {this} implies from \eqref{sl: 7806}, that whenever $ -m_0 > m_1$ we expect $h^0 = 0$. The green boundary, however, appears later, at $-2m_0 = m_1$, so {the vanishing of the slope only provides} a lower bound for the green boundary. A similar lower bound can be derived from the slope for the occurrence of the green region in the other examples.
	
	Moreover, we can also study the line bundle index over the manifold 7806, which is given in \eqref{eq: 7806i}. In the red region in the first quadrant, the index has to be positive. Continuing counter clockwise, when  passing the green line to the purple region we have $h^0 = 0$ and $h^1 \neq 0$, thus at some point the index has to turn negative. Studying the change of sign of the index over the second quadrant, we find, that it tends to flip around the curve
	  \[m_0 = -\frac{m_1^3+4m_1}{3m_1^2+2} \; .
	  \]
	   Hence, we find an upper bound for when the green line can appear. It is also possible to use the index to find upper bounds for the green regions on the other manifolds. We conclude that the vanishing loci of the slope and index provide approximate boundaries for the green region.

	\subsection{$h^1$ features}
	
	As a consequence of Kodaira vanishing theorem we have that the first and third quadrant of every $h^1$ plot has to be zero. Following the results of section \ref{sec:311} we find that for the ambient space $\mathcal{A} = \mathbb{P}^1 \times \mathbb{P}^4$ the fourth quadrant has to be zero. For $\mathcal{A} = \mathbb{P}^2 \times \mathbb{P}^3$, this is not necessarily the case as trivial $h^1$ here translates to the vanishing of a kernel in \eqref{eq: 234h1}. We will come back to this later.
	
	Starting from the first quadrant and going counterclockwise we encounter a blue region with polynomial description 
	\begin{align}
	h^1 =  m_0 (c_1 - c_2m_0^2).
	\end{align}
	This polynomial guarantees a smooth transition from the zero region at $m_0 = 0$, where both descriptions overlap. The blue region is bounded from below by the green line/region, which we discussed in the previous section. It is very interesting, that in this region $h^1$ only depends on the line bundle charge coming from the ambient projective space with lowest dimension.
	
	Below the green region we encounter in red the region where $h^1$ is given by the index. In the case of $\mathcal{A} = \mathbb{P}^1 \times \mathbb{P}^4$ the red region stretches to $m_1 = 0$. This is expected, since we found here $h^0=0$ and we know from section \ref{sec:311} that $h^2 = 0 = h^3$. Our results show, that the image of the map in \eqref{eq: 142h1} is never onto, which by Serre duality is equivalent to stating that the kernel in \eqref{eq: 144h2} does not vanish. On the yellow line at $m_1 = 0$, we have
	\begin{align}
		h^1 = -1 - m_0 .
	\end{align}
	
	On the other hand for $\mathcal{A} = \mathbb{P}^2 \times \mathbb{P}^3$ we also encounter a red region below the green line, when the kernel in \eqref{eq: 232h0} vanishes. The occurrence of the red region is interesting in itself, since it implies that for these values the map in \eqref{eq: 232h2} is onto and $h^2 = 0$. The red region is bounded below by the purple one. This boundary must occur whenever the index becomes positive, so that it has to be compensated by non zero $h^2$ values. In fact, we find the non trivial result, that whenever the index becomes positive 1) the contribution of the maps in \eqref{eq: 232h1} cancel and 2) the map in \eqref{eq: 232h2} is no longer onto. By Serre duality this translates to \eqref{eq: 234h1} having a non trivial kernel and the cancellation of maps in \eqref{eq: 234h2}.
	
	However, for the manifolds in the set $G_2$ we found black points at the boundary between the red and purple regions. At these points we have various non-trivial and non-maximal maps so that $h^1 \neq 0 \neq h^2$. This is only possible for the manifolds in set $G_2$, not for those in $G_1$ and $G_3$, cf. \eqref{eq: 142h0}  and \eqref{eq: 232h0}. In principle we should have expected many more such points, or possibly even its own region, where all maps are generic. Our results, however, show that black points only occur when the sign of the index is close to flipping. Solving for a change in sign of the index, in fact, shows that the Serre dual lines found in the previous section will only approximately hold for the full range of line bundle charges. We thus expect more black points for larger charges, and possibly shifts in the exact location of the lines that these points lie on; what remains certain is that these special points will lie near the locus where the index changes sign.
	
\begin{figure}
		\begin{minipage}{0.50\textwidth}
			\centering
			\includegraphics[width=1.1\textwidth]{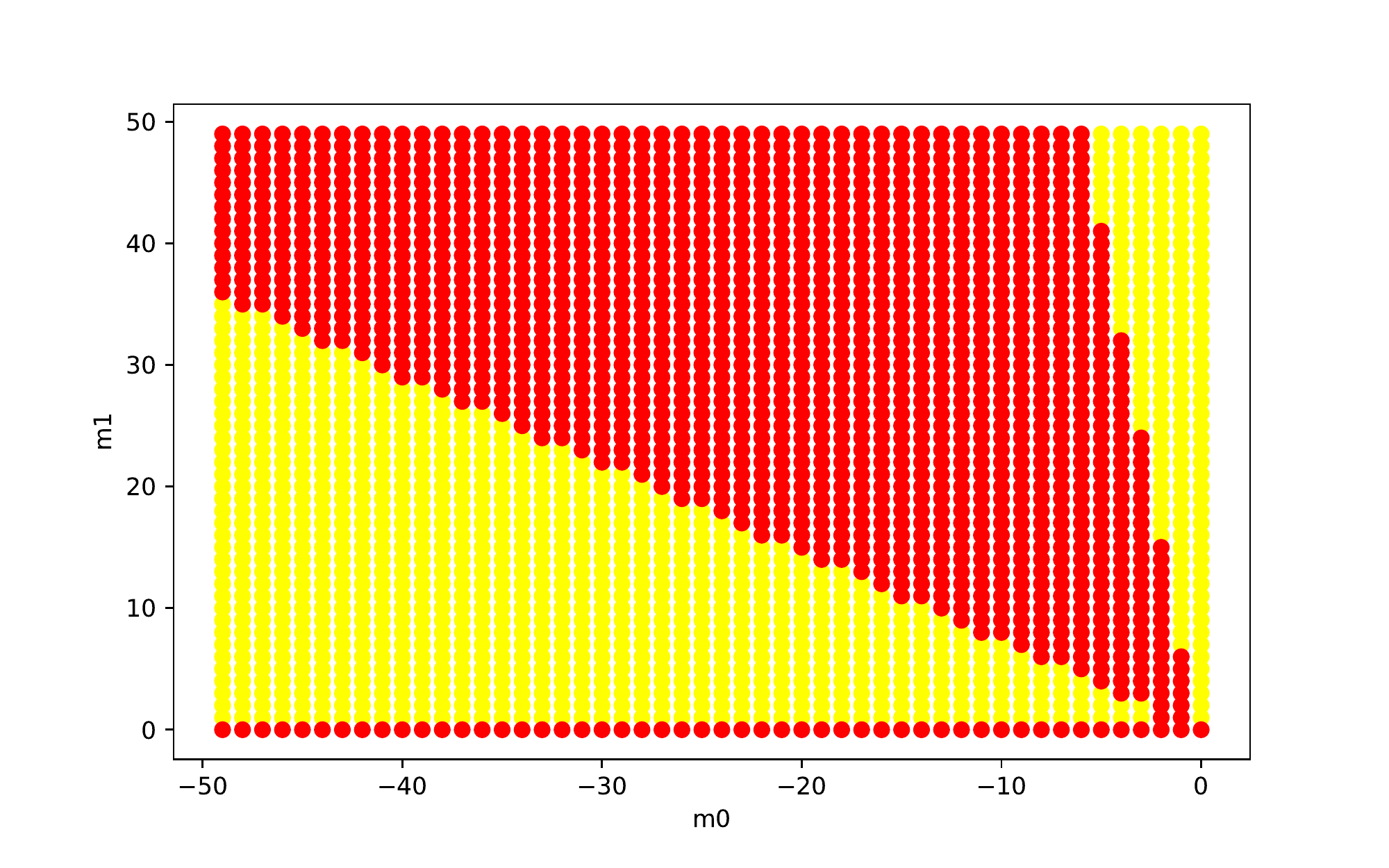}
		\end{minipage}\hfill
		\begin{minipage}{0.50\textwidth}
			\centering
			\includegraphics[width=1.1\textwidth]{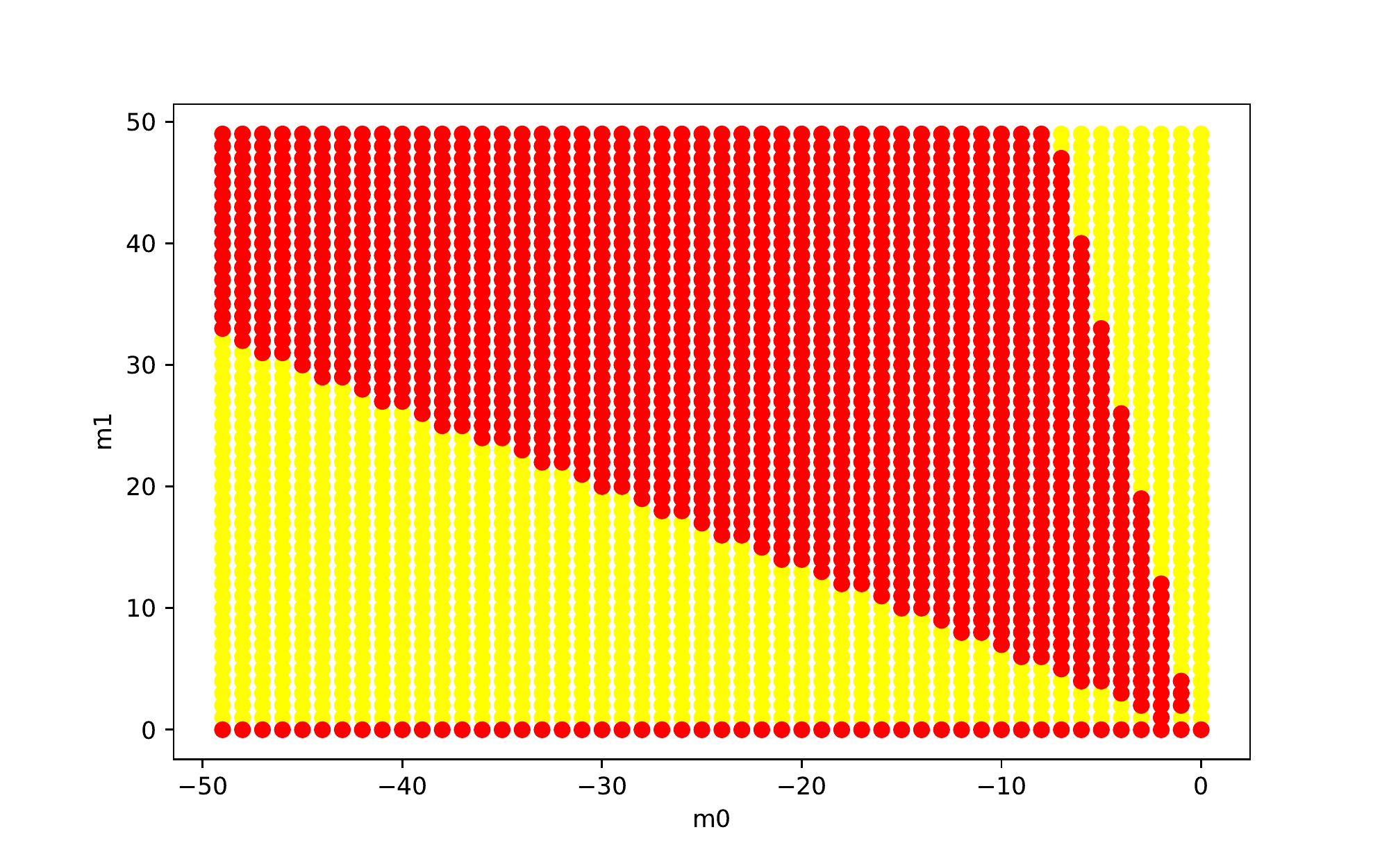} 
		\end{minipage}
		\caption{Plots of the change in sign of the indices \eqref{in: 7844} (left) and \eqref{in: 7883} (right) for the charges in the second quadrant. Red (yellow) indicates that the index is negative (positive).} 
		\label{plt: flip}
	\end{figure}

	Let us determine whether the change in index sign is a good indicator of where the polynomial expressions break down, so that black points result. For the manifolds where we observed this behaviour,  the indices of the line bundles, \eqref{in: 7833}, \eqref{in: 7844} and \eqref{in: 7883}, are all quadratic in $m_0$ and of degree three in $m_1$. Further, the quadratic term in $m_0$ always comes with a $m_1$. As a consequence,  the flip in index sign will follow at leading order $m_0 \propto m_1$.  Take the index of $M_{7883}$ and solve for $\ind(L) = 0$
	\begin{align}
		m_0 = -\frac{5}{6} m_1 \pm \frac{1}{6} \sqrt{17 m_1^2 - 88}
	\end{align}
	which yields two lines. Taking the negative sign, we find at leading order  $m_0 \approx - \frac{9.12}{6} m_1 \approx - \frac{3}{2} m_1$ approximately the Serre dual line found in \eqref{eq: 7883h1}. The positive solution for $m_0$ describes a line somewhere in the blue region of the plots in \ref{plt: 7883b}. A visualization of this results is shown in figure \ref{plt: flip}.
	Similarly, we can find for $M_{7844}$ from \eqref{in: 7844}
	\begin{align}
		m_0 = - \frac{9}{12} \left(m_1 + \frac{1}{m_1} \right) \mp \frac{1}{12} \sqrt{57m_1^2 -102 +\frac{81}{m_1^2} } 
	\end{align} 
	which at leading order becomes $m_0 \approx - \frac{16.55}{12} m_1 \approx - \frac{4}{3} m_1$ reproducing approximately the results found in \eqref{eq: 7844h1}. Figure \ref{plt: flip} shows the change of sign in \eqref{in: 7844} for the second quadrant. We can see, that our earlier result $m_0 = -\frac{4}{3} m_1$ works quite well, but fails for some values. Take for example the charges (-33,24) according to our classification we have $h^1=0$, however in \ref{plt: flip} we see that the index turns negative,
	\begin{align}
		\ind(\mathcal{O}(-33,24)) = -155
	\end{align}
	and thus $h^1 > h^2$ here. An analysis of the index \eqref{in: 7833} again shows an approximate match of our results found in \eqref{eq: 7833h1} and the leading orders.
	
	When $h^1 = 0$ and the index becomes positive, we need to have $h^2 \neq 0$ in the second quadrant. This is reflected by Serre duality in a red region in the fourth quadrant. The implications of this result on the maps have been discussed above. 
	The red region is then as usual bounded by the yellow line. It sits on the boundary between fourth and first quadrant at $m_1 = 0$ and corresponds to the polynomial description
	\begin{align}
	h^1 = \frac{1}{2} (m_0 -1) (m_0 -2).
	\end{align}

	\section{Outlook}
	
	\label{sec:5}
	
	In this paper, we have analysed line bundle cohomologies on the eight favourable co-dimension two CICY manifolds of Picard number two. Our analysis extends and supports the observations of Refs \cite{Constantin:2018hvl,Klaewer:2018sfl}. We find that the variation of the cohomology data with respect to the line bundle charges partitions the line bundle charge space into several regions. For most cases, these regions can be described as cones starting at the origin. We further find  analytical expressions of the cohomology dimensions in  these regions. In the majority of cases, these are given by simple polynomials of degree three in the line bundle charges, but in one specific example we were only able to derive a recursive description of the  data. We also observed in some examples, that a few points lying on the  boundary between regions defy an analytical description. Such special points are an exception, and only appear for three of the studied manifolds, and we showed that they can be predicted by studying the charges for which the index of the line bundle changes sign. 	
	
	The line bundle cohomologies share many similarities over the eight different examples considered. In particular, the general form of the polynomials, and  the partitioning of the line bundle charge space into regions, agree to a remarkable degree between the manifolds. 
	These similarities between different Calabi Yau manifolds suggest that it could be possible to derive the cohomology dimensions, by studying similar, or possibly even simpler, manifolds. For example the polynomials and regions of
	\begin{align}
	M_{7887} =  \left[
	\begin{array}{c||c}
	1 & 2 \\
	3 & 4  
	\end{array}
	\right]^{2,86}_{-168} \nonumber
	\end{align}
	described in \cite{Constantin:2018hvl} follow the same pattern and form as found in this paper.
	This observation mimics a feature of the original CICY list, namely that new CICY manifolds can be generated via a splitting mechanism from simple ones. To further explore the role played by the splitting mechanism for line bundle cohomologies, it would be interesting to investigate CICYs with Picard number three that result from splitting of the manifolds studied here.

	Another promising approach for the study of line bundle cohomologies might be to use techniques from data science, such as in \cite{Ruehle:2017mzq} where line bundle cohomologies were predicted using genetic algorithms. It has also been shown that machine learning algorithms yield solid results in predicting the Hodge numbers, fibrations and discrete symmetries over the whole CICY list \cite{Bull:2018uow,He:2017aed,He:2019vsj}. Moreover, machine learning has been successfully used in the study of CY hypersurfaces in toric ambient spaces \cite{Klaewer:2018sfl}. Finally, in a more recent study \cite{Halverson:2019tkf} it was demonstrated, although in a different geometric setting, that these algorithms show some remarkable flexibility. That might just be whats needed to derive predictions for line bundle cohomologies on CICY manifolds related by splitting.  
	
	Our work provides one more piece of information for line bundle cohomologies on CY manifolds, but many questions remain. In particular, there are two outstanding questions that would complete the analytic approach advocated in this paper.  A first question, is if we can predict the exact position of the green and blue regions, without extensive algebro-geometric computations? The results we present provides support for such a prediction, based on analysis of the slope and index of the line bundle.
	More importantly, can we predict the exact coefficients of the polynomials (or recursive relations) that describe line bundle cohomologies on CICY manifolds? Answering this question in the affirmative would provide a very powerful tool to phenomenological studies of string theory, and gives a clear motivation for further studies of this topic.
	
	\bigskip
	As this paper was prepared for publication, we were informed by the upcoming publications \cite{brodie19}, where line bundle cohomologie formulas for surfaces are derived, and found to follow a pattern similar to the Calabi--Yau manifolds studied in this paper.

	\section*{Acknowledgments}
	
	We are grateful to M. Magill for fruitful discussions and  L. Anderson, A. Constantin, J. Gray and A. Lukas for providing very valuable feedback regarding the computation of line bundle cohomologies. RS would like to thank A. Wehrhahn and G. K\"alin for help with various computer issues. RS would like to thank L. Schlechter for some clarification regarding earlier stages of this project. 
	Data creation was carried out on resources provided by the Swedish National Infrastructure for Computing (SNIC) at Uppsala Multidisciplinary Center for Advanced Computational Science (UPPMAX).
	M.L. and R.S are financed by  the Swedish Research Council (VR) under grant numbers 2016-03873 and 2016-03503.
	
	\appendix

\bibliographystyle{JHEP}

\bibliography{ref}

\end{document}